\documentclass[12pt]{article}
\usepackage[utf8]{inputenc}
\usepackage{amsthm,amsmath,amssymb}
\usepackage[margin=.7in]{geometry}
\usepackage{setspace}
\usepackage{graphicx}
\usepackage{subcaption,placeins}
\usepackage{colortbl}
\usepackage{ragged2e}
\usepackage{hyperref}
\hypersetup{
    colorlinks=true,
    urlcolor=blue,
    citecolor=blue,
    linkcolor=blue
}
\PassOptionsToPackage{hyphens}{url}\usepackage{hyperref}
\PassOptionsToPackage{sloppy}{url}\usepackage{hyperref}
\usepackage{cleveref}
\usepackage{ctable}
\usepackage{indentfirst}
\usepackage{makecell}
\usepackage{booktabs}
\usepackage{multirow}

\newcommand{\E}{\mathbb{E}}
\renewcommand{\P}{\textrm{P}}

\usepackage[bibstyle=numeric, citestyle=authoryear, doi=false, url=true, backend=biber, maxbibnames=50, maxcitenames=4, uniquelist=false, uniquename=false, sorting=nyt]{biblatex}
\renewbibmacro{in:}{}
\DeclareNameAlias{default}{family-given/given-family}
\newcommand{\citet}{\textcite}
\newtheorem{assumption}{Assumption}
\newtheorem{proposition}{Proposition}
\newtheorem{lemma}{Lemma}
\newtheorem{theorem}{Theorem}
\newtheorem{remark}{Remark}
\newtheorem{inneruassumption}{}
\newenvironment{namedassumption}[1]
  {\renewcommand\theinneruassumption{#1}\inneruassumption}
  {\endinneruassumption}

\crefname{assumption}{Assumption}{Assumptions}
\crefname{lemma}{Lemma}{Lemmas}

\title{Policy Evaluation during a Pandemic\thanks{We thank the editor, associate editor, and two anonymous referees for their insightful comments.  We also thank Andrew Goodman-Bacon, Laura Hatfield, Jonathan Roth, Ian Schmutte, Meghan Skira, and seminar participants at Brown University and the 2021 Annual Health Econometrics Workshop for a number of helpful comments.}}

\author{Brantly Callaway\thanks{Department of Economics.  University of Georgia.  \href{mailto:brantly.callaway@uga.edu}{brantly.callaway@uga.edu}} \and Tong Li\thanks{Department of Economics.  Vanderbilt University.  \href{mailto:tong.li@vanderbilt.edu}{tong.li@vanderbilt.edu}}}

\newcommand\independent{\protect\mathpalette{\protect\independenT}{\perp}}
    \def\independenT#1#2{\mathrel{\setbox0\hbox{$#1#2$}%
    \copy0\kern-\wd0\mkern4mu\box0}} 

\begin{document}

\maketitle

\abstract{National and local governments have implemented a large number of policies in response to the Covid-19 pandemic.  Evaluating the effects of these policies, both on the number of Covid-19 cases as well as on other economic outcomes is a key ingredient for policymakers to be able to determine which policies are most effective as well as the relative costs and benefits of particular policies.  In this paper, we consider the relative merits of common identification strategies that exploit variation in the timing of policies across different locations by checking whether the identification strategies are compatible with leading epidemic models in the epidemiology literature.  We argue that unconfoundedness type approaches, that condition on the pre-treatment ``state'' of the pandemic,  are likely to be more useful for evaluating policies than difference-in-differences type approaches due to the highly nonlinear spread of cases during a pandemic.  For difference-in-differences, we further show that a version of this problem continues to exist even when one is interested in understanding the effect of a policy on other economic outcomes when those outcomes also depend on the number of Covid-19 cases.  We propose alternative approaches that are able to circumvent these issues.  We apply our proposed approach to study the effect of state level shelter-in-place orders early in the pandemic.}

\vspace{40pt}


\bigskip


\vspace{80pt}

\pagebreak

\onehalfspacing

\section{Introduction}

There have been a large number of policies implemented in order to decrease the spread of Covid-19.  During the early part of the pandemic, the most important of these policies were non-pharmaceutical interventions such as requirements to wear masks, making Covid-19 tests widely available, contact tracing, school closures, lockdowns, and others.  These sorts of policies are likely to come with a number of tradeoffs in terms of effectiveness in reducing the spread of Covid-19 as well as their effects on individuals' economic and psychological well-being.  Thus, understanding the effects of different policies along a number of dimensions (both effects on number of cases as well as effects on other outcomes) is a key ingredient for researchers, policymakers, and governments to consider when evaluating Covid-19 related policies.

The main way that these policies have been studied by researchers is to compare outcomes in locations that implemented some policy to outcomes in another location that did not implement the policy.  Researchers typically exploit having access to panel data --- data on cases, testing, and economic variables is generally widely available for particular locations over multiple time periods --- to try to understand these effects.  This sort of setup is very familiar to many researchers in economics, and, with this sort of data availability, an almost default strategy of empirical researchers is to use difference-in-differences.  And, indeed, difference-in-differences has been widely used to study the effects of policies in response to Covid-19.

In the current paper, we argue that difference-in-differences has properties that make it relatively less attractive for conducting policy evaluation during a pandemic than it typically would be for most applications in economics.  The intuition for our results is that difference-in-differences methods are typically motivated by a two-way fixed effects model for untreated potential outcomes (see, for example, \citet{blundell-dias-2009}).  The key feature of these models is that they include additively separable unit-level unobserved heterogeneity (i.e., a fixed effect).  This sort of heterogeneity is very common in applications in economics; a textbook example would be an application on the effect of some policy on individuals' earnings where the researcher is worried that ``ability'' is unobserved, affects earnings, and is distributed differently between the group of individuals that are affected by the policy and the group of individuals not affected by the policy.  In this setup, the unobserved heterogeneity can be differenced out and paths of outcomes for the group of treated units and the group of untreated units can be compared to each other to deliver the effect of the policy.

However, this sort of motivation does not apply in the case of Covid-19.  In particular, the main epidemiological models for Covid-19 transmission are highly nonlinear and depend on (i) the number of currently infected individuals in a particular location, (ii) the number of susceptible individuals in a location, and (iii) the transmission properties of Covid-19.  In other words, the key challenge for identifying effects of Covid-19 related policies on the number of Covid-19 cases is not that different locations are different in terms of unobserved heterogeneity, but rather but that pre-policy differences in the ``state of the pandemic'' between treated and untreated locations (e.g., differences in the number of Covid-19 cases before the policy is implemented) can lead to substantial differences between locations in how the pandemic would have evolved absent the policy being implemented.  These differences can make it difficult to evaluate the effects of Covid-19 related policies.  Moreover, these differences are a major concern for evaluating early pandemic policies where different locations' policy choices often depended on the state of the pandemic in those locations.%

Another central issue in the economics literature is to understand the effect of Covid-19 related policies on various economic outcomes.\footnote{Throughout the text, we use the term ``economic outcomes'' but our results apply to any outcome of interest that is outside the epidemic model that we consider in the paper.}  Understanding the effects of Covid-19 related policies on economic outcomes is essential in order to understand the costs and benefits of various policies that are aimed at reducing the number of Covid-19 cases.  For economics outcomes, we focus on the simple leading case where untreated potential outcomes are generated by a two way fixed effects model that also depends on the current number of Covid-19 cases in a particular location.  This setup allows both for the active number of Covid-19 cases to have an effect on the outcome of interest and for the active number of cases to themselves be affected by the policy.   In this case, we consider (i) a ``standard'' version of difference-in-differences that directly compares paths of economic outcomes among treated and untreated locations, and (ii) difference-in-differences when the current number of Covid-19 cases is included as a regressor.  We show that, generally, neither approach can deliver the average effect of the treatment on economic outcomes across treated locations.  The first strategy breaks down because of differences in the current number of cases across treated and untreated locations.  The second strategy breaks down when the policy affects the number of Covid-19 cases (which is the goal of the policy).

We propose alternative approaches that address both sets of issues mentioned above.  In particular, for evaluating the effect of policies on Covid-19 cases, we first show that unconfoundedness-type identification strategies (i.e., strategies that compare locations that have the same pre-treatment values of key Covid-19 related variables) do not suffer from the same drawbacks as difference-in-differences approaches.  For this case, we propose an estimation strategy that involves estimating (i) the propensity score (i.e., the probability of experiencing the policy conditional on pre-treatment values of Covid-19 related variables) and (ii) an outcome regression for Covid-19 cases in the absence of the policy that is related to the epidemic model.  Our approach is doubly robust in the sense that it delivers consistent estimates of policy effects if either the propensity score or outcome regression is correctly specified.  This is important as it implies that we can circumvent having to estimate a full structural epidemic model while still delivering estimates of policy effects that are \textit{compatible} with the epidemic model.  %

Second, for evaluating the effect of Covid-19 related policies on economic outcomes, we propose a two-step approach where the parameters of a two way fixed effects model that additionally allows for current cases to affect the outcome are identified using untreated locations in the first step.  Then, in the second step, we recover the path of active cases that treated locations would have experienced on average if they had not been treated (this follows under similar unconfoundedness type arguments used for cumulative cases above).  Using these two pieces of information, we are able to construct the average economic outcome that treated locations would have experienced if they had not participated in the policy --- and, therefore, the average effect of the policy is identified for treated locations.  Unlike other common approaches, this approach allows both for the policy to affect the current number of cases and for the current number of cases to affect untreated potential outcomes.

We conclude the paper by studying the effect of state-level shelter-in-place orders (SIPOs) on the number of Covid-19 cases and on travel early in the pandemic.  These are challenging policies to evaluate because states that implemented these policies also tended to have a large number of Covid-19 cases earlier than states that did not implement this type of policy (or implemented it later).  This correlation mechanically leads to larger increases in Covid-19 cases among early treated states relative to untreated and later-treated states.  This additionally implies that parallel trends is violated and can lead to difference-in-differences estimates that these policies \textit{increased} the number of Covid-19 cases which is clearly unreasonable.  We additionally show that difference-in-differences estimates are very sensitive to minor changes in how they are specified.  And, interestingly, despite very different post-policy estimates across different types of DID specifications, none of them are rejected in pre-treatment periods.  This suggests that it is not feasible to choose between alternative difference-in-differences specifications based on their performance in pre-treatment periods.  Using difference-in-differences, we also sometimes spuriously estimate meaningfully large effects of placebo SIPOs in states that did not actually implement a SIPO.  In contrast, %
we document notably better performance of the unconfoundedness strategy along several dimensions: this approach does not lead to any estimates that SIPOs increased Covid-19 cases, and it does not lead to large or statistically significant effects of placebo SIPOs among states that did not actually implement a SIPO.   %

\subsubsection*{Related Work}

There are a number of recent papers at the intersection of economics, Covid-19, and policy evaluation, and here we only briefly summarize some of the most related ones.  

The most related papers to ours are several methodological papers on evaluating Covid-19 related policies.   \citet{allcott-boxell-conway-ferguson-gentzkow-goldman-2020} propose an event-study regression estimator that is motivated by SIRD models (the same type of model that we consider below) though their approach ends up being substantially different from ours.  \citet{goodman-marcus-2020,gauthier-2021} discuss two-way fixed effects regressions in the context of evaluating Covid-19 policies.  \citet{chernozhukov-kasahara-schrimpf-2021} propose an alternative approach to evaluate Covid-19 related policies that is motivated by a SIRD model.  Their approach is more structural than ours which comes with tradeoffs.  For example, their approach can be useful for evaluating counterfactual policies while ours is geared towards evaluating the effects of policies that were actually implemented.  On the other hand, our approach generally requires fewer assumptions to evaluate policies that were actually enacted and is set up to be robust to general forms of treatment effect heterogeneity (which is likely to be important in contexts like many early pandemic policies where policies were often implemented at different times in different locations; see \citet{goodman-marcus-2020} for some related discussion).  \citet{aleman-busch-ludwig-santaeulalia-2020} provide a way to transform different pandemic ``states'' across locations and evaluate pandemic-related policies exploiting different locations being in different ``stages''.

Difference-in-differences has been widely used in empirical work to study the effects of Covid-19 policies.  Some examples of papers that consider Covid-19 related policies using difference-in-differences types of identification strategies include \citet{bartik-bertrand-lin-rothstein-unrath-2020}, \citet{berry-fowler-glazer-handel-macmillen-2021}, \citet{chetty-friedman-hendren-stepner-2020}, \citet{courtemanche-garuccio-le-pinkston-yelowitz-2020}, \citet{dave-friedson-matsuzawa-mcnichols-sabia-2020}, \citet{dave-friedson-matsuzawa-sabia-safford-2020}, \citet{gapen-millar-blerina-sriram-2020}, \citet{glaeser-jin-leyden-luca-2021}, \citet{goolsbee-syverson-2021}, \citet{gupta-montenovo-nguyen-rojas-schmutte-simon-weinburg-2020}, \citet{haynes-kulkarnia-li-siddique-2022}, \citet{hsiang-et-al-2020},  \citet{juranek-zoutman-2021}, \citet{kong-prinz-2020}, \citet{villas-sears-villas-villas-2020},  \citet{wright-sonin-driscoll-wilson-2020}, and \citet{ziedan-simon-wing-2020}.  \citet{haber-et-al-2022} provides a recent survey of empirical strategies that have been used to evaluate Covid-19 related policies.

Finally, on the econometrics side, our paper is related to a large literature on unconfoundedness and difference-in-differences (see \citet{imbens-wooldridge-2009} for a survey of this literature).  More notably, our results on checking the compatibility of structural epidemic models with reduced form identification strategies is broadly similar to a number of papers in econometrics; for example, just in the context of panel data, \citet{heckman-robb-1985,heckman-ichimura-todd-1997,athey-imbens-2006,blundell-dias-2009, chabe-2015,ghanem-santanna-wuthrich-2022,marx-tamer-tang-2022} all provide connections between structural models and conditions under which various reduced form approaches, such as difference-in-differences, can be compatible with these models.  Our contributions on allowing for infections to both be affected by the policy and to have a direct effect on economic outcomes appear to be conceptually new but related to work on mediation analysis (see \citet{huber-2020} for a summary of this literature) which is also broadly related to work on simultaneous equation models (e.g., \citet{griliches-1977,imbens-newey-2009}).  Viewing the ``untreated potential'' number of Covid-19 infections as a covariate in the model for untreated potential outcomes is related to the idea of difference-in-differences with covariates that can be affected by the treatment; see, \citet{bonhomme-sauder-2011,lechner-2011,caetano-callaway-payne-rodrigues-2022} for related discussion along these lines.

\section{Motivating Examples}

\begin{figure}[t!]
    \centering
    \caption{A Simulated SIRD Model}
    \label{fig:sim-sird}
    \includegraphics[width=.7\textwidth]{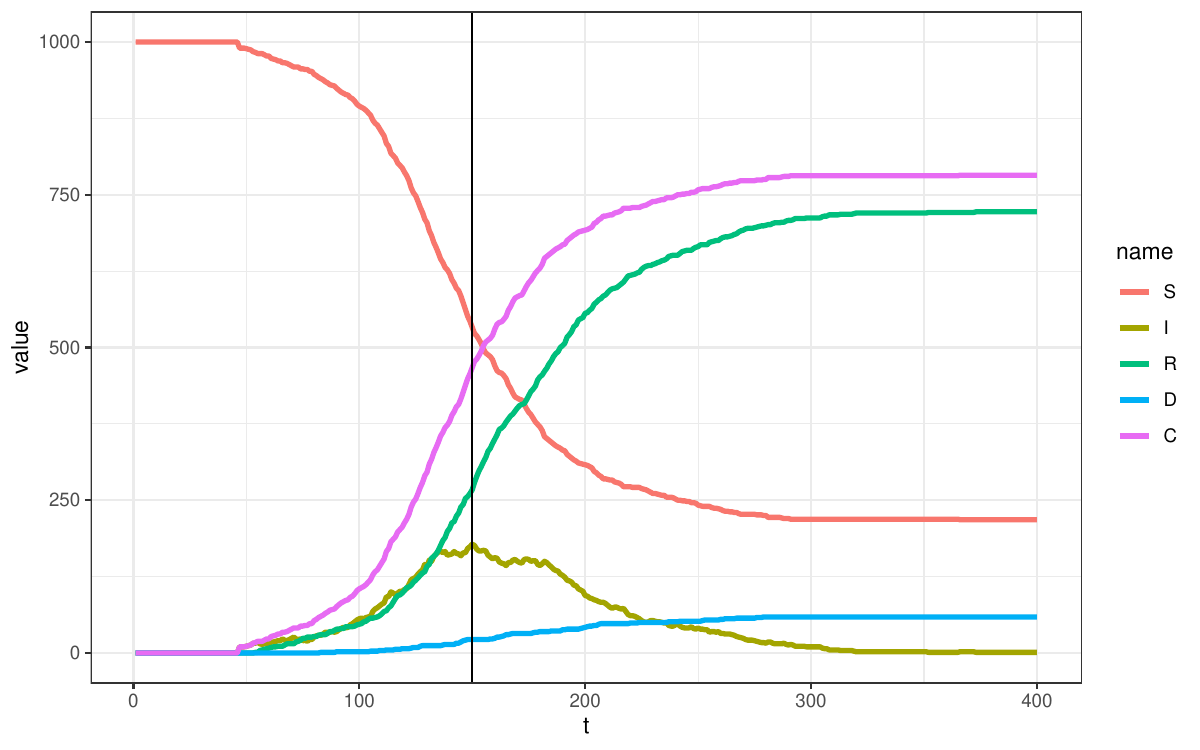}
    
    \begin{justify}
    {\footnotesize \textit{Notes:} The figure shows paths of each variable in a stochastic SIRD model for one (treated) location in the simulated data.  The vertical black line is placed at $t=150$ when the policy was implemented.  $S$ stands for the number of susceptible individuals in the population, $I$ stands for the number of currently infected individuals, $R$ stands for the total number of recovered individuals in the population, $D$ stands for the cumulative number of deaths, and $C$ stands for the cumulative number of cases.  In this example, the total population is 1000, there are 400 time periods, and the policy has no effect on the pandemic.  The specific values for the parameters in this simulation are provided in \Cref{tab:sim-params} in \Cref{app:sim-details}.}
    \end{justify}
\end{figure}

To start with, in this section we provide examples of the main types of issues that can confound policy analysis during a pandemic using simulated data from the leading type of epidemic model that has been widely used in the context of Covid-19.  \Cref{fig:sim-sird} shows the paths of the key variables during a simulated pandemic coming from a stochastic SIRD model. SIRD models categorize individuals in a population into being S-Susceptible, I-Infected, R-Recovered, or D-Dead.  We discuss this model in substantially more detail in the next section.  The shapes of the path of each variable is typical of a SIRD model.  In particular, at some point in time, some small number of cases shows up in a particular location.  Then, the number of infections rise in early periods when there are a large number of susceptible individuals in that location combined with an increasing number of currently infected (which also implies contagious).  As the number of susceptible decreases (i.e., as infected individuals recover or die), eventually the number of infected individuals decreases.  Simultaneously, the cumulative number of cases, number of recovered individuals, and number of deaths all initially grow before eventually leveling off.

\begin{figure}[t!]
    \centering
    \caption{Simulated Policy Effects on Cumulative Cases}
    \label{fig:sim-policy-effects}
    \begin{subfigure}[b]{.49\textwidth}
    \centering \includegraphics[width=\textwidth]{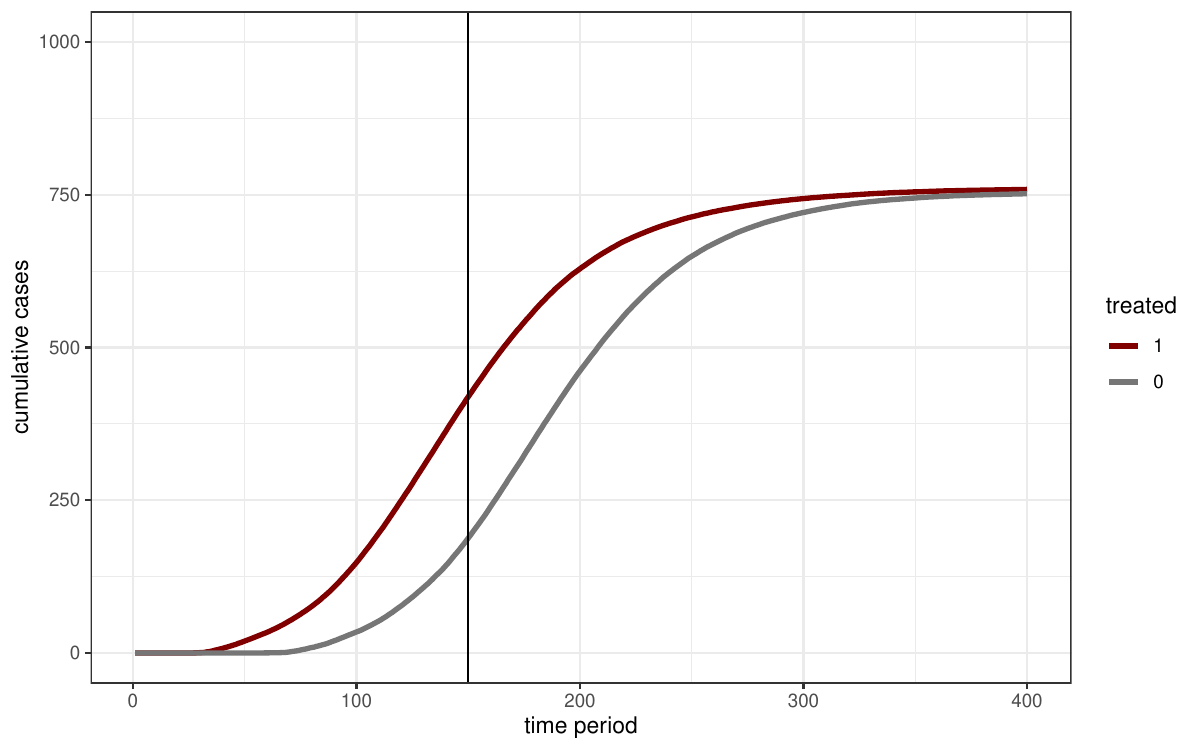}
    \caption{Paths of Cumulative Cases}
    \end{subfigure}
    \begin{subfigure}[b]{.49\textwidth}
    \centering \includegraphics[width=\textwidth]{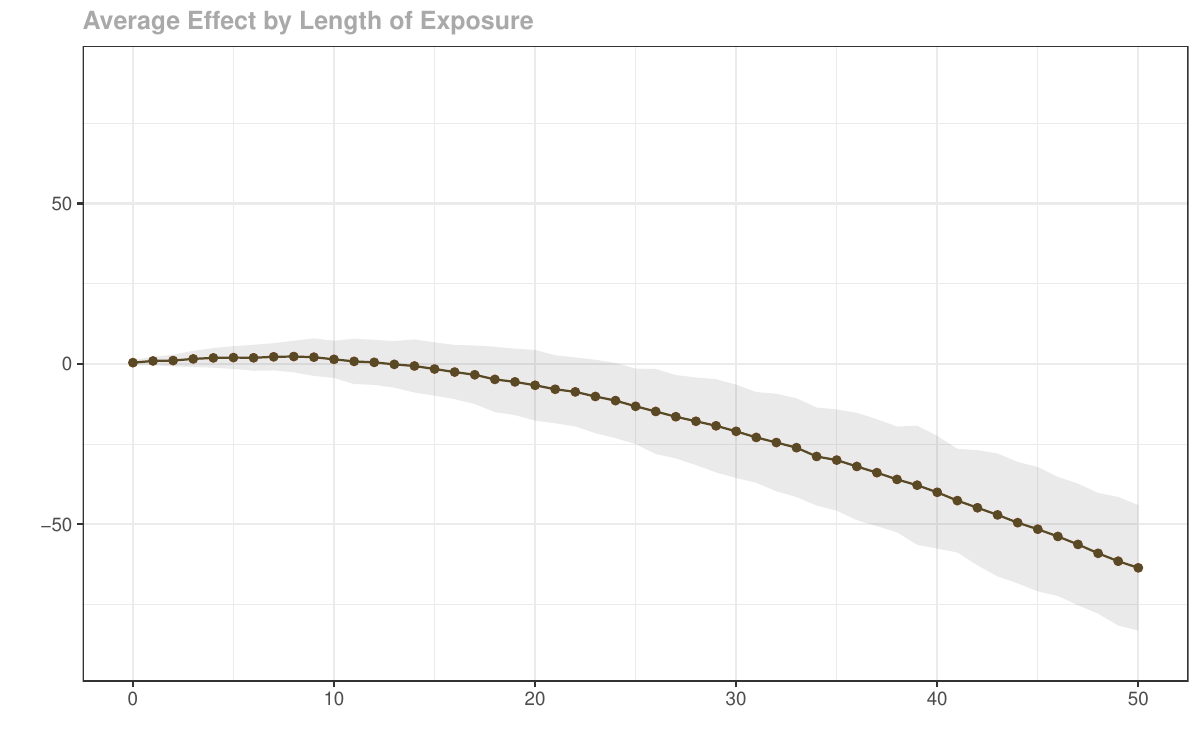}
    \caption{DID Event Study Estimates}
    \end{subfigure}
    \begin{justify}
    { \footnotesize \textit{Notes:} Panel (a) plots simulated average paths of cumulative cases among treated and untreated locations.  The vertical black line is at $t=150$ when the policy is implemented for the treated locations.  In this example, the policy is constructed so that it has no effect of Covid-19 cases.  Panel (b) plots event study type estimates based on a parallel trends assumption for the effect of the policy on the number of cases by length of exposure to the treatment.}
    \end{justify}
\end{figure}

In this example, we consider the case where a new policy is implemented in some locations in period 150.  For simplicity, we consider the case where the policy has no effect on Covid-19 cases.  Locations that participate in the treatment and locations that do not participate in the treatment are alike in all ways except that treated locations tend to experience their first cases earlier than untreated locations.  Panel (a) of \Cref{fig:sim-policy-effects} shows plots of the average paths of cumulative cases for treated locations and untreated locations in this setup.  

Panel (b) of \Cref{fig:sim-policy-effects} shows event study-type estimates of the effect of the treatment on the number of cases.  To be precise, these are difference-in-differences type estimates where the estimated effect comes from the average change in cases experienced by the treated group of locations relative to the change in cases experienced by the untreated group of locations over the same time periods.  Taken at face value, the estimated effects in Panel (b) suggest that the policy decreased the number of Covid-19 cases in treated locations relative to what they would have been in the absence of the policy.  However, recall that, in our simulation setup, the policy has no effect on Covid-19 cases.  Thus, this example demonstrates that difference-in-differences can perform poorly in the context of trying to evaluate the effect of a policy on the number of Covid-19 cases.  The key driver of this poor performance is (i) the nonlinearity of the model for Covid-19 transmission and (ii) differences in the timing of the first cases between locations that participate in the treatment and those that do not.  The first of these is an inherent feature of trying to evaluate the effects of policies on Covid-19 cases.  For the latter, generally, the bias of difference-in-differences approaches for policy evaluation becomes more severe as the timing of first cases becomes more different between treated and untreated locations.\footnote{Interestingly, the best case for difference-in-differences is when the timing of first cases is the same across treated and untreated locations.  However, this is also a case where there is no need to take a time difference at all and one could just make level comparisons of Covid-19 cases across locations.}  

\begin{figure}[t!]
    \centering
    \caption{Simulated Policy Effects on Economic Outcome}
    \label{fig:sim-economic-outcomes}
    \begin{subfigure}[b]{.49\textwidth}
    \centering \includegraphics[width=\textwidth]{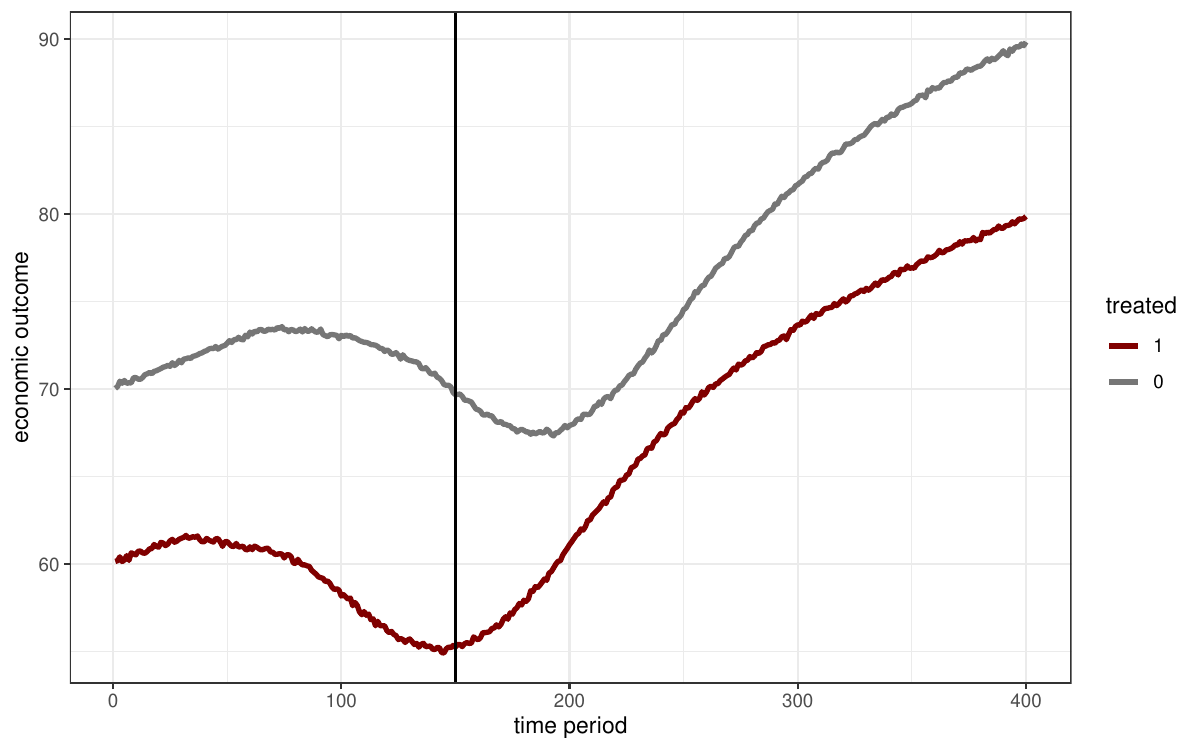}
    \caption{Paths of Economic Outcome}
    \end{subfigure}
    \begin{subfigure}[b]{.49\textwidth}
    \centering \includegraphics[width=\textwidth]{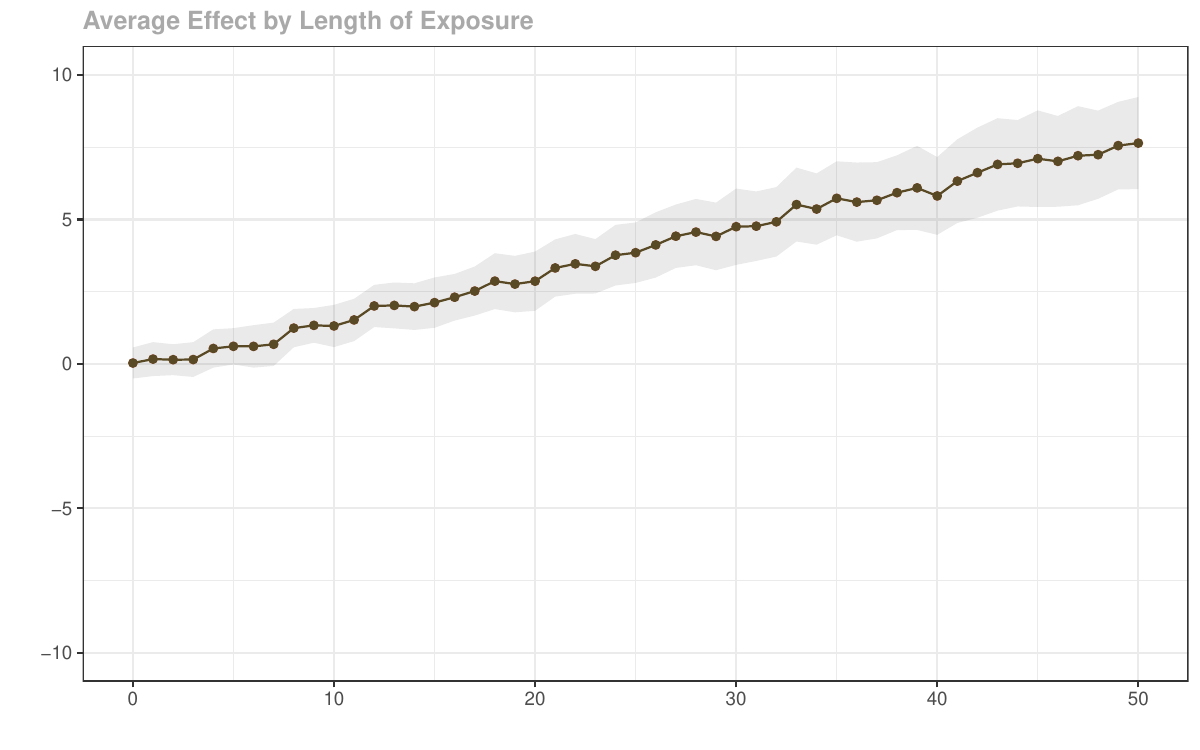}
    \caption{Event Study Estimates}
    \end{subfigure}
    \begin{justify}
    { \footnotesize \textit{Notes:} Panel (a) plots simulated average paths of outcomes among treated and untreated locations.  The vertical black line is at $t=150$ when the policy is implemented for treated locations.  In this example, the policy is constructed so that it has no effect on economic outcomes (either directly or through its effect on Covid-19 cases).  Panel (b) plots event study type estimates based on a parallel trends assumption for the effect of the policy on the number of cases by length of exposure to the treatment.}
    \end{justify}
\end{figure}

Next, we consider the effect of the policy on some economic outcome of interest.  \Cref{fig:sim-economic-outcomes} continues with the same simulated policy as above.  As above, we consider the case where the policy has no effect on cases or on economic outcomes.  However, in this simulation we allow for the economic outcome to depend on the number of active Covid-19 cases in a particular location (here, more active cases tend to decrease the economic outcome), but, otherwise, the economic outcome would follow parallel trends.  Panel (a) shows average paths of outcomes for treated and untreated locations in this setup.  Panel (b) shows event study type estimates under the assumption of parallel trends.  As before, and even in this very simple example, differences in the timing of first cases lead to violations of parallel trends that lead to poor estimates of the effect of the policy on the economic outcome of interest.

\begin{figure}[t!]
    \centering
    \caption{Estimated Policy Effects under Unconfoundedness}
    \label{fig:sim-unc}
    \begin{subfigure}[b]{.49\textwidth}
    \centering \includegraphics[width=\textwidth]{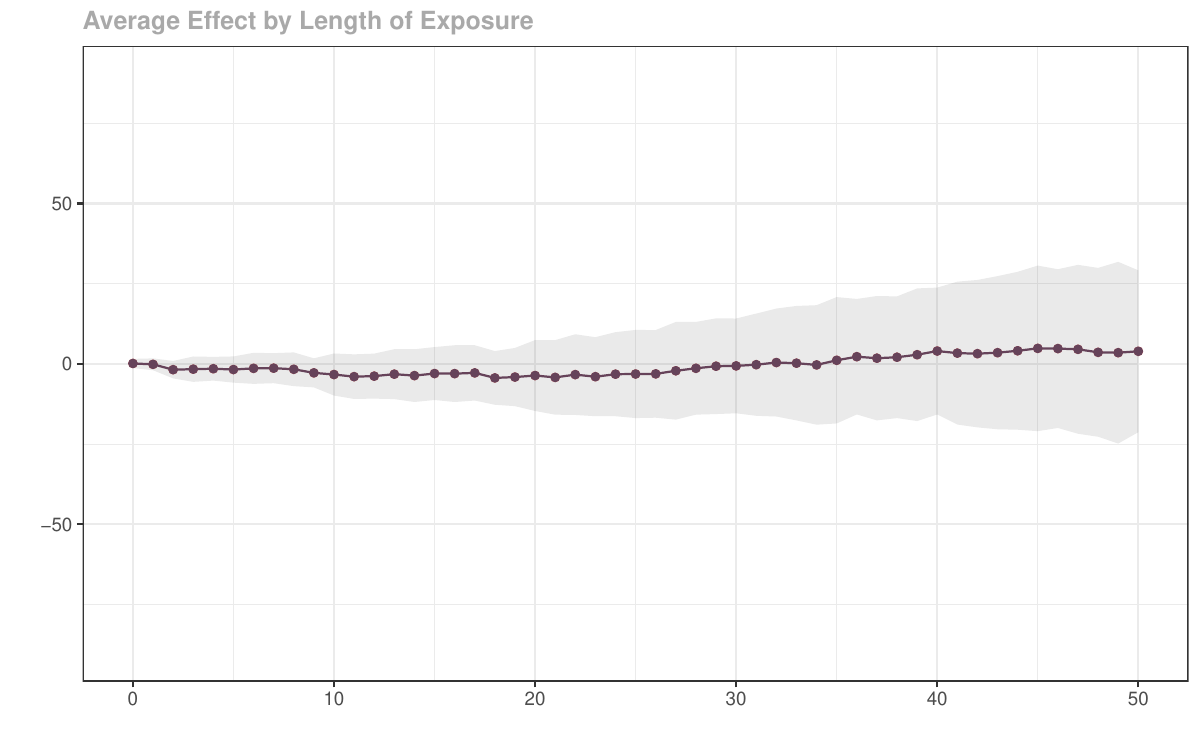}
    \caption{Event Study for Cumulative Cases}
    \end{subfigure}
    \begin{subfigure}[b]{.49\textwidth}
    \centering \includegraphics[width=\textwidth]{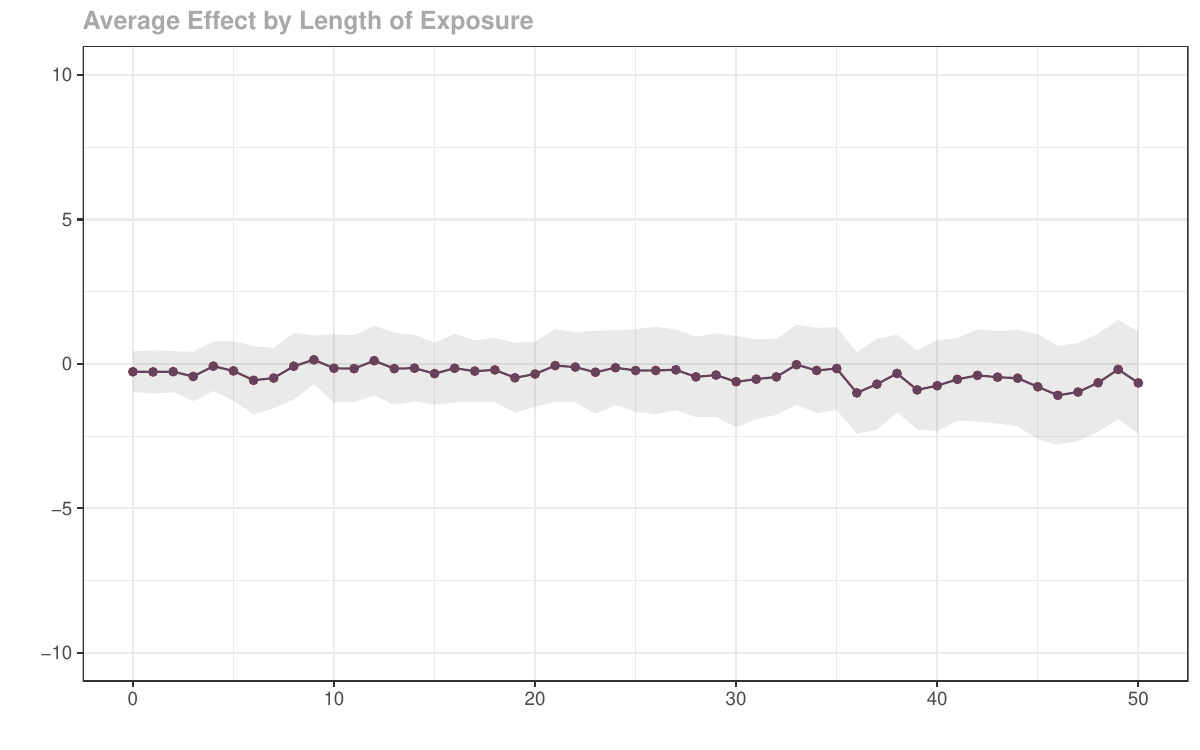}
    \caption{Event Study for Economic Outcome}
    \end{subfigure}
    \begin{justify}
    { \footnotesize \textit{Notes:} This figure plots event study-type estimates of the effect of a simulated policy on cumulative cases (in Panel (a)) and on an economic outcome (in Panel (b)) using the unconfoundedness-type identification arguments considered in the paper and using the doubly robust estimator discussed below.  In this simulation, the policy has no effect on either outcome.  }
    \end{justify}
\end{figure}

Finally, we contrast the poor performance of difference-in-differences in both of these contexts with using an unconfoundedness type strategy to deal with the pandemic-related variables.  In particular, when Covid-19 cases is the outcome, we effectively compare locations that were in a similar pandemic ``state'' in the period right before the policy was implemented.  For the economic outcome, we continue to use a version of difference-in-differences, but one that, in the absence of the policy, accounts for economic outcomes depending on the number of Covid-19 infections that would have occurred if the policy had not been implemented using an unconfoundedness strategy (see \Cref{sec:economic-outcomes} below for more details).  Estimates using these approaches are provided \Cref{fig:sim-unc}.  In both cases, these strategies perform notably better at evaluating the effects of the policy.

\section{A Baseline Stochastic SIRD Model}

In this section, we briefly discuss a stochastic SIRD model which is the workhorse model of epidemic spread in epidemiology and has been used extensively to forecast the spread of Covid-19 cases.  SIRD models have a long history in epidemiology --- a deterministic version of this kind of model was proposed by \citet{kermack-mckendrick-1927}.  Stochastic SIRD models are discussed in \citet{allen-2008,allen-2017} and have been considered by economists in \citet{oka-wei-zhu-2021, fernandez-jones-2022,ellison-2020,acemoglu-chernozhukov-werning-whinston-2021}, among  others.

\paragraph{Notation:}  Let $N_l$ denote the number of individuals in location $l$.  Let $\mathcal{T}$ denote the total number of time periods.  The number of susceptible individuals in location $l$ in a particular time period $t$ is denoted by $S_{lt}$, the number of currently infected individuals in location $l$ at time period $t$ is denoted by $I_{lt}$, the cumulative number of recovered individuals is denoted by $R_{lt}$, and the number of cumulative deaths is denoted by $\delta_{lt}$.  All individuals in the population are in exactly one of these states at a particular point in time so that
\begin{align*}
    N_l = S_{lt} + I_{lt} + R_{lt} + \delta_{lt}
\end{align*}
in all time periods.  Later, we will be interested in the effect of the policy on the cumulative number of cases by time period $t$, and we denote this variable by $C_{lt}$ and note that $C_{lt} = N_l - S_{lt}$.

In a SIRD model, the paths of all of these variables are governed by some transition equations.  The transition equations have the Markov property; i.e., the path of each outcome over time only depends on the ``state'' of location $l$ in the immediately preceding period.   And, in particular, these transition equations are given by 
\begin{align} 
    \E[I_{lt} | \mathcal{F}_{lt-1}] &= (1 - \lambda - \gamma) I_{lt-1} + \beta \frac{I_{lt-1}}{N_l} S_{lt-1} \label{eqn:transition-infections} \\
    \E[R_{lt} | \mathcal{F}_{lt-1} ] &= R_{lt-1} + \lambda I_{lt-1} \label{eqn:transition-recover} \\
    \E[\delta_{lt} | \mathcal{F}_{lt-1} ] &= \delta_{lt-1} + \gamma I_{lt-1} \label{eqn:transition-deaths} \\
    \E[S_{lt} | \mathcal{F}_{lt-1}] &= S_{lt-1} - \beta \frac{I_{lt-1}}{N_l} S_{lt-1} \label{eqn:transition-susceptible} \\
    \E[C_{lt} | \mathcal{F}_{lt-1}] &= C_{lt-1} + \beta \frac{I_{lt-1}}{N_l}S_{lt-1} \label{eqn:transition-total-cases}
\end{align}
where for some time period $t$, we define $\mathcal{F}_{lt} = (S_{lt},I_{lt},R_{lt},\delta_{lt})$, and often refer to this as the ``state'' of the pandemic in location $l$ in time period $t$.  It is worth considering each of these equations in some more detail.  To start with, consider the term $\beta \frac{I_{lt-1}}{N_l}S_{lt-1}$ which shows up in \Cref{eqn:transition-infections,eqn:transition-susceptible,eqn:transition-total-cases}.  This is the expected number of new cases in time period $t$ conditional on the state of the pandemic in location $l$ in time period $t-1$.  The expected number of new cases from one period to the next depends on three things.  First, it depends on $I_{lt-1}/N_l$ which is the fraction of individuals that are infected in period $t-1$.  Holding other things constant, when more individuals are infected, it implies an expected larger increase in the number of cases.  Second, the expected number of new cases depends on the number of susceptible individuals in the population.  Intuitively, when there are more susceptible individuals, the number of cases grows more rapidly (other things constant).  The spread of a pandemic stops when the number of susceptible becomes small enough which can happen either through ``herd immunity'' or by decreasing the number of susceptible (for example, through the introduction of a vaccine).  Finally, the change in the number of cases depends on the parameter $\beta$ which is called the infection rate.  Most non-pharmaceutical interventions are aimed at changing the infection rate --- here, there are two potential benefits: (i) decreasing the infection rate through non-pharmaceutical interventions decreases the total number of cases that need to occur before reaching herd immunity,\footnote{This is also one explanation for repeated ``waves'' of Covid-19 cases.  That is, the infection rate may be temporarily reduced by policy intervention or individual choices but then increases again once these interventions are relaxed.} and (ii) if there is a vaccine on the horizon, it also would decrease the total number of cases that occur before herd immunity is reached through the vaccine.  

Next, consider \Cref{eqn:transition-recover}.  This transition equation says that, on average, the number of total recoveries in location $l$ in time period $t$ (conditional on the state of the pandemic in period $t-1$) is equal to number of individuals in location $l$ that have already recovered by time period $t-1$ plus some fraction of infected individuals in period $t-1$.  This fraction is determined by the parameter $\lambda$ which is the recovery rate from Covid-19.  \Cref{eqn:transition-deaths} is the transition equation for deaths.  The key parameter is $\gamma$ which parameterizes the death rate from being infected with Covid-19.  

Next, consider \Cref{eqn:transition-infections}.  This is the transition equation for active Covid-19 cases.  The expected number of infections in period $t$ thus depends on (i) the remaining cases after accounting for recoveries and deaths (this is the first term in \Cref{eqn:transition-infections}), and (ii) the expected number of new cases (this is the second term in \Cref{eqn:transition-infections}).  Finally, in \Cref{eqn:transition-susceptible}, the expected number of susceptible individuals is equal to the number of susceptible individuals in time period $t-1$ minus the expected number of new cases; likewise, the expected number of cumulative cases by time period $t$ is equal to the number of cumulative cases in time period $t-1$ plus the expected number of new cases.

\section{Identification Strategies for Policy Effects on Covid\mbox{-}19 Cases} \label{sec:cum-cases}

The previous section presented a basic stochastic SIRD model.  This section connects that sort of model with the treatment effects literature and considers the relative merits of difference-in-differences and unconfoundedness strategies for evaluating the effect of a policy on the number of Covid-19 cases.  

The strategy of this section is to impose the stochastic SIRD model for untreated potential outcomes and to check if difference-in-differences and/or unconfoundedness are compatible with the stochastic SIRD model.  This setup does not place restrictions on how treated potential outcomes (i.e., Covid-19 cases under the policy) are generated.  In particular, this is consistent with Covid-19 cases under the policy continuing to follow a stochastic SIRD model but where the values of the parameters potentially change in response to the policy; but it is also more general than that in the sense that there are no substantive restrictions on treated potential outcomes.  Perhaps more importantly, this setup also allows for heterogeneous effects of policies across different locations.

\paragraph{Additional Treatment Effects Notation} To make the connection with the treatment effects literature, we start by introducing some additional notation.  First, we define $D_l$ as a binary variable indicating whether or not location $l$ participated in the treatment.  We also define treated and untreated versions of all of the variables in the stochastic SIRD model.  In particular, for generic time period $t$, $S_{lt}(0)$, $I_{lt}(0)$, $R_{lt}(0)$, and $\delta_{lt}(0)$ are the number of susceptible, infected, recovered, and dead individuals in location $l$ in time period $t$ if the policy had not been enacted.  We also define $C_{lt}(0)$ as the cumulative number of cases in location $l$ by time period $t$ if the policy had not been enacted. Similarly, we define $S_{lt}(1)$, $I_{lt}(1)$, $R_{lt}(1)$, $\delta_{lt}(1)$, and $C_{lt}(1)$ to be the corresponding treated potential variables; i.e., the values of each of these if the policy had been enacted.  Following a large literature on policy evaluation which exploits having access to panel data, we consider the case where the researcher has access to some pre-treatment periods.  We suppose that the policy is implemented for treated locations in time period $t^*$ where $1 < t^* \leq \mathcal{T}$.\footnote{In practice, the timing of implementing a particular policy may vary across different locations.  Extending our arguments to this case is relatively straightforward, and, therefore, this section considers the case where the policy is implemented at the same time across all treated locations.  See \Cref{rem:attgt} below for additional discussion on this point.} For random variables indexed by time periods, we define $\Delta X_t := (X_t - X_{t-1})$.  Because we are also interested in how policy effects vary across time, some of our arguments involve ``long differences'' where, for $t_2 > t_1$, we define $\Delta^{(t_1,t_2)} X_t := X_{t_2} - X_{t_1}$.  In \Cref{sec:sird-for-untreated}, we write the SIRD model given in the previous section in terms of untreated potential outcomes, and we refer to this model as the  \Cref{ass:sird-model} throughout the remainder of the paper.

Our main interest for this part of the paper is the effect of the policy on the cumulative number of Covid-19 cases.  Typically, the main parameter of interest in DID applications (and the parameter that we focus on in the current paper) is the Average Treatment Effect on the Treated (ATT).  It is given by
\begin{align} \label{eqn:attc}
    ATT^C_t = \E[C_t(1) - C_t(0) | D=1]
\end{align}
where we index the $ATT$ by $C$ to indicate that we are considering the effect of the policy on the cumulative number of cases in time period $t$.  $ATT^C_t$ is the difference between cumulative cases under the policy relative to cumulative cases in the absence of the policy on average among locations that participated in the treatment.  That this parameter is disaggregated by time period makes it straightforward to report across time periods (as in an event study), but it is also straightforward to, for example, average it across post-treatment time periods in order to report an overall average effect of participating in the treatment.

\subsection{Using Difference-in-Differences to Evaluate Policy Effects on Covid-19 Cases}

The main underlying motivation for considering a DID approach is when a researcher thinks that untreated potential outcomes are generated from a two-way fixed effects model (see, for example, \citet{blundell-dias-2009}).  These sorts of models are attractive in many applications in economics where there are thought to be important unobserved differences between individuals (or firms, etc.) that are not observed by the researcher.  In labor economics, these are often thought of as being unobserved skill; in industrial organization, these may be unobserved differences in productivity across firms; and, in health economics, these may be thought of as proneness to particular health conditions.  However, there is an important difference of Covid-19 relative to all of these cases.  In general, particular locations do not have time invariant unobservables that make them more or less likely to have a large number of cases; instead, the key differences between locations are (i) the timing of their initial case(s), and (ii) the pandemic response (both in terms of policies and in terms of actions taken by the populations in different locations).\footnote{One caveat to this is that different locations may have characteristics that are related to the parameters of the SIRD model discussed above.  See \Cref{rem:model-extensions} below for more discussion along these lines.}

The main result in this section is that there are likely to be major drawbacks to using DID to evaluate the effects of Covid-19 related policies on the number of Covid-19 cases.  The two primary reasons for this are (i) the highly nonlinear spread of Covid-19 cases during a pandemic and (ii) that the key difference between locations is the current number of Covid-19 cases rather than some fixed unobserved difference between locations in terms of ``proneness'' to having a large number of cases.  

In this section, we consider whether difference-in-differences approaches are compatible with the stochastic SIRD model presented above.  We begin by providing some background on using difference-in-differences to identify the effect of some policy.  The key identifying assumption in a DID application is the following parallel trends assumption.

\begin{namedassumption}{Parallel Trends Assumption} \label{ass:parallel-trends} For all $t=2,\ldots,\mathcal{T}$
\begin{align*}
    \E[\Delta C_t(0) | D=1] = \E[\Delta C_t(0) | D=0]
\end{align*}
\end{namedassumption}
The parallel trends assumption says that the path of Covid-19 cases that locations in the treated group would have experienced if they had not participated in the treatment is the same as the path of Covid-19 cases that locations in the untreated group did experience.  Invoking this assumption leads to the following estimand for the $ATT^C_t$ for $t \geq t^*$
\begin{align} \label{eqn:att-estimand}
    DID^C_t = \E[\Delta^{(t^*-1,t)} C_t | D=1] - \E[\Delta^{(t^*-1,t)} C_t | D=0]
\end{align}
where we use the notation $DID^C_t$ to highlight that it may not be equal to $ATT^C_t$.  $DID^C_t$ is equal to the path of Covid-19 cases that treated locations experienced adjusted by the path of Covid-19 cases that untreated location experienced; if the parallel trends assumption holds, then the latter is the path of Covid-19 cases that treated locations would have experienced on average if they had not experienced the policy, and $DID^C_t$ would be equal to $ATT^C_t$.  And, regardless of whether or not the parallel trends assumption holds, $DID^C_t$ is the population quantity for what is estimated in DID applications on Covid-19.  Before providing our main result on using difference-in-differences to identify/estimate the effect of a policy on Covid-19 cases, it is also worth mentioning that the primary motivating model for difference-in-differences identification strategies is one where 
\begin{align} \label{eqn:twfe}
    C_{lt}(0) = \theta_t + \eta_l + v_{lt}
\end{align}
where $\theta_t$ is a time fixed effect, $\eta_l$ is location-specific unobserved heterogeneity that can be distributed differently between the treated group and untreated group and $v_{lt}$ is an idiosyncratic time varying unobservable.  Comparing \Cref{eqn:twfe} to the equation for cumulative Covid-19 cases in \Cref{eqn:transition-total-cases}, it is immediately clear that these are notably different.  In the stochastic SIRD model, the important difference between treated and untreated locations is not unobserved heterogeneity, but rather differences in the current number of Covid-19 cases and the number of susceptible individuals across locations.  This immediately provides a suggestive piece of evidence that the parallel trends assumption is unlikely to hold when $C_{lt}(0)$ is generated from a stochastic SIRD model.  

The next result makes explicit that the parallel trends assumption is generally violated in stochastic SIRD models and provides an expression for the bias resulting from incorrectly imposing the parallel trends assumption in cases where untreated potential outcomes are generated by a stochastic SIRD model.
\begin{theorem}  \label{thm:did-bias} In the stochastic SIRD model discussed above
\begin{align*}
    \E[\Delta^{(t^*-1,t)} C_t(0) | D=d] =  \sum_{s=t^*}^{t} \E\Big[ \E[\Delta C_s(0) | \mathcal{F}_{t^*-1}, D=d] | D=d\Big]
\end{align*}
which implies that
\begin{itemize}
    \item[(i)] The parallel trends assumption does not generally hold
    \item[(ii)] Further, the bias from incorrectly imposing the parallel trends assumption is given by
    \begin{align*}
        & \left( \E[\Delta^{(t^*-1,t)} C_t | D=1] - \E[\Delta^{(t^*-1,t)} C_t | D=0] \right) - ATT^C_t \\
        & \hspace{25pt} = \left(\sum_{s=t^*}^{t} \E\Big[ \E[\Delta C_s(0) | \mathcal{F}_{t^*-1}, D=1] | D=1\Big] - \sum_{s=t^*}^{t} \E\Big[ \E[\Delta C_s(0) | \mathcal{F}_{t^*-1}, D=0] | D=0\Big]\right)
    \end{align*}
\end{itemize}
\end{theorem}

The proof of \Cref{thm:did-bias} is provided in \Cref{sec:proofs}.  \Cref{thm:did-bias} shows that difference-in-differences generally delivers (potentially severely) biased estimates of the effect of a policy on cumulative Covid-19 cases.  It is worth making a few additional comments before proceeding.  First, the key reason why the difference-in-differences strategy breaks down is that, in general, the distribution of pandemic related variables immediately before the policy (contained in $\mathcal{F}_{t^*-1}$) is not the same across treated and untreated locations.  Due to the nonlinearity of the SIRD model, this leads to violations of the parallel trends assumption. Second, the sign of the bias cannot generally be determined from these expressions.  For example, in \Cref{fig:sim-policy-effects} above, difference-in-differences resulted in downward biased estimates of the effect of the policy, but the direction of the bias is sensitive to both (i) timing of first cases in treated and untreated locations, and (ii) the timing of the policy itself (this can be clearly seen in Panel (a) of \Cref{fig:sim-policy-effects} where setting the policy at an alternative time period could result in parallel trends being violated in the opposite direction).  

Some of the expressions in \Cref{thm:did-bias} seem complicated. One special case of this result that is worth pointing out is when $t=t^*$ (so that we are considering the effect of the policy on Covid-19 cases ``on impact'').  In that case, the bias from using DID is given by 
\begin{align*}
    \E\left[\beta \frac{I_{t^*-1}(0)}{N} S_{t^*-1}(0) \big| D=1\right] - \E\left[\beta \frac{I_{t^*-1}(0)}{N} S_{t^*-1}(0) \big| D=0\right]
\end{align*}
This bias is the difference between the expected number of new cases that treated locations would have experienced in the absence of the policy relative to the expected number of new cases for untreated locations.  And, here, it is straightforward to see key reasons why difference-in-differences can perform poorly: if the joint distribution of currently infected and number of susceptible individuals is different among treated and untreated locations, then they would have experienced a different number of new Covid-19 cases \textit{even if the policy had not been implemented}.  In the context of Covid-19, there are some cases where these biases could be substantial.  Perhaps the leading example is when the timing of initial Covid-19 cases varied across locations and Covid-19 related policies were implemented earlier in locations that tended to have cases earlier.

\subsection{Unconfoundedness in SIRD Models}

A main alternative to difference-in-differences for evaluating the effects of policies is to assume some version of unconfoundedness.  Unconfoundedness means that, after conditioning on some covariates, treatment assignment is as good as randomly assigned.  In other words, in order to identify the effect of some policy on Covid-19 cases, one can compare Covid-19 cases in locations that experienced the treatment to Covid-19 cases in locations that did not participate in the treatment \textit{and had the same characteristics related to the pandemic} as treated locations.  In this section, we consider a particular version of unconfoundedness that does not suffer from the same limitations as difference-in-differences for evaluating the effects of policies on the number of Covid-19 cases.  

Intuitively, the reason why an unconfoundedness strategy works better for studying policy effects of Covid-19 is that the key differences between locations are the current amount of cases and the current number of susceptible individuals rather than differences in location-specific unobserved heterogeneity.  Therefore, conditioning on current cases and the current number of susceptible individuals is sufficient for comparisons of treated and untreated locations to deliver causal effects of policies on Covid-19 cases; while the differencing strategy of difference-in-differences is not able to do the same.

The next result is a main result on the validity of identifying policy effects under the assumption of unconfoundedness.  Before stating this result, define the propensity score as 
\begin{align*}
    p(\mathcal{F}_{t^*-1}) := \P(D=1 | \mathcal{F}_{t^*-1})
\end{align*}
which is the probability of being treated conditional on pre-treatment characteristics $\mathcal{F}_{t^*-1}$ and make the following assumption
\begin{assumption}[Overlap] \label{ass:overlap} There exists some $\epsilon > 0$ such that $\P(D=1) > \epsilon$ and $p(\mathcal{F}_{t^*-1}) < 1 - \epsilon$ almost surely.
\end{assumption}
\Cref{ass:overlap} is a standard assumption in the treatment effects literature.  In the context of Covid-19 related policies, the first part says that there are some locations that participate in the treatment, and the second part says that, for all values of $\mathcal{F}_{t^*-1}$, one can find untreated locations that have those characteristics.  This implies that, for all treated locations, there exists matching untreated locations with the same pre-treatment characteristics.  In practice, if this condition is violated, one can identify treatment effects that are local to the region of common support (see, for example, \citet{crump-hotz-imbens-mitnik-2009}).

\begin{proposition} \label{prop:unc} In the \Cref{ass:sird-model} and under \Cref{ass:overlap}, and for any $t \geq t^*$,  
\begin{align*}
    \E[C_{t}(0) | \mathcal{F}_{t^*-1}, D=1] = \E[C_{t}(0) | \mathcal{F}_{t^*-1}, D=0]
\end{align*}
\end{proposition}
The proof of \Cref{prop:unc} is provided in \Cref{sec:proofs}.  This is an important result and implies that, on average, the unobserved number of cumulative cases that locations that participated in the treatment would have experienced if they had not participated in the treatment is the same as the cumulative number of cases that untreated locations actually did experience \textit{among locations that had the same pre-treatment characteristics.}

Finally, in this section, we provide an identification result for $ATT^C_t$ which is valid under the SIRD model for Covid-19 cases.    

\begin{theorem}\label{thm:unc} In the \Cref{ass:sird-model} and under \Cref{ass:overlap}, and for any $t \geq t^*$,  
\begin{align} \label{eqn:att-dr}
    ATT^C_t =\E\left[ \omega(D,\mathcal{F}_{t^*-1}) \left( C_{t} - m^C_{0,t}(\mathcal{F}_{t^*-1}) \right)  \right]
\end{align}
where
\begin{align} \label{eqn:weights}
    \omega(D,\mathcal{F}_{t^*-1}) := \frac{D}{\E[D]} - \frac{\frac{p(\mathcal{F}_{t^*-1})}{1-p(\mathcal{F}_{t^*-1})} (1-D)}{\E\left[\frac{p(\mathcal{F}_{t^*-1})}{1-p(\mathcal{F}_{t^*-1})} (1-D)\right]} \qquad \textrm{and} \qquad m^C_{0,t}(\mathcal{F}_{t^*-1}) := \E[C_t|\mathcal{F}_{t^*-1},D=0]
\end{align}
\end{theorem}

\Cref{thm:unc} says that, under a stochastic SIRD model, we can evaluate the effect of a policy using an unconfoundedness strategy that compares the number of cases in locations that participated in the treatment to the number of cases in locations that did not participate in the treatment \textit{and} which had the same Covid-19 related characteristics in the period before the policy was implemented.

It is worth making several additional comments related to the result in \Cref{thm:unc}.  First, estimating $ATT^C_t$ from the expression in \Cref{eqn:att-dr} involves estimating the propensity score, $p(\mathcal{F}_{t^*-1})$ and the outcome regression $m^C_{0,t}(\mathcal{F}_{t^*-1})$.  It is also possible to derive alternative expressions for $ATT^C_t$ that only require either estimating the propensity score (these would be similar to propensity score re-weighting estimators as in \citet{hirano-imbens-ridder-2003}) or estimating the outcome regression (these would be similar to regression adjustment estimators).  However, the expression for $ATT^C_t$ in \Cref{eqn:att-dr} possesses the double robustness property.\footnote{For completeness, we provide a proof in the Supplementary Appendix, but the arguments follow along the same lines as arguments for existing doubly robust estimators under unconfoundedness.}  A main advantage of a doubly robust estimator is that it provides consistent estimates of $ATT^C_t$ if either the propensity score model or the outcome regression model are correctly specified (see, for example, \citet{bang-robins-2005,sloczynski-wooldridge-2018}).  Double robustness is particularly appealing in this context as it enables us to side-step the problem of estimating the full SIRD model and instead involves estimating a model of the treatment assignment process which is both familiar to economists and may be substantially more feasible to do with a simple parametric model.  In unreported simulations, we found that imposing flexible parametric models for both the propensity score and the outcome regression performed notably better than either the pure outcome regression approach or the propensity score re-weighting approach.  

Second, it is worth briefly mentioning that the weights in \Cref{eqn:att-estimand} are normalized to have mean one in finite samples.  This type of normalized weights is said to be of the H{\'a}jek-type (\citet{hajek-1971}) and typically results in estimators with improved finite sample properties relative to its unnormalized counterpart (\citet{busso-dinardo-mccrary-2014}). Finally, we provide the asymptotic properties of our estimator in the Supplementary Appendix.  In order to conduct inference, we use a multiplier bootstrap procedure that involves perturbing the influence function of the estimator of $ATT^C_t$; we also discuss how to conduct uniform inference across different time periods to account for multiple testing.  These results primarily follow from recent results on doubly robust estimators with H{\'a}jek-type weights in \citet{santanna-zhao-2020}.

\bigskip

\bigskip

\begin{remark}
The results in this section have focused on the effect of a policy on the number of cumulative cases.  However, the same sorts of arguments apply to other possible variables of interest such as the current number of infections.  For example, the proof of \Cref{prop:unc} additionally covers the other variables in the epidemic model, and we use similar arguments as in \Cref{thm:unc} but for the current number of infections in the next section.
\end{remark}

\begin{remark} \label{rem:attgt}
 The identification arguments in this section have been for the case where the timing of the policy does not vary across different locations.  However, it is straightforward to extend these arguments to the case where the timing of the policy varies across locations (as is the case in our application).  In this case, one can think of our identification arguments holding specifically for each ``group'' where a group is defined by the time period when a location first becomes treated.  In this case, instead of identifying ATT-type parameters, one would identify group-time average treatment effects along the lines of \citet{callaway-santanna-2021} and can follow their approach to aggregating these sorts of parameters into an overall average effect of participating in the treatment or into an event study type result.  We provide a complete discussion of this extension in the Supplementary Appendix.  There is variation in treatment timing in the application that we consider below, and this is the approach that we follow in the application.
\end{remark}

\begin{remark} \label{rem:model-extensions}
    In the Supplementary Appendix, we consider a number of extensions to the \ref{ass:sird-model} and pay particular attention to which sorts of extensions are or are not compatible with the unconfoundedness condition discussed in this section.  The types of extensions that we consider allow for the parameters of the SIRD model to vary by location and time periods; for example, instead of the infection rate $\beta$ being constant across locations and time periods, we instead denote the infection rate by $\beta_{lt}$ and consider what sort of additional structure on $\beta_{lt}$ is compatible with unconfoundedness.  First, we show that unconfoundedness is generally compatible with SIRD model parameters varying over time; e.g., $\beta_{lt}=\beta_t$.  This can be empirically relevant if, for example early in the pandemic, the availability of masks changed over time or when there is common time-varying information about how Covid-19 is transmitted.  On the other hand, we also show that the unconfoundedness approach is not compatible with model parameters that are specific to particular locations.  One example is when $\beta_{lt} = \exp(\theta_t + \theta_l)$ where $\exp(\cdot)$ enforces that the infection rate should be positive and $\theta_t$ is a time fixed effect and $\theta_l$ is a location fixed effect in the infection rate.  An empirically relevant exception to this limitation is when the location-specific fixed effects can be accounted for by observed location-specific characteristics.  For example, arguably the most important reasons for location-specific variation in infection rates are likely to be differences in population density, age and/or income distribution, and demographic characteristics --- all of which are generally observed (see, \citet{chernozhukov-kasahara-schrimpf-2021} for some related discussion along these lines).
\end{remark}

\section{Identification Strategies for Policy Effects on Economic Outcomes}

\label{sec:economic-outcomes}

Another interest of economists is studying the effect of Covid-19 related policies on other (particularly economic) outcomes.  This is likely to be useful for thinking about a cost-benefit analysis of particular policies.  Relative to textbook versions of difference-in-differences, what is different in this section is that we allow for economic outcomes to depend on the current number of Covid-19 cases in a particular location.\footnote{As above, because the target parameter is an ATT-type parameter, the setup in this section does not require assumptions on how treated potential outcomes are generated and, therefore, the discussion about the effect of current cases and SIRD models in this section need only apply for untreated potential outcomes.}
We denote the economic outcome of interest by $Y_{lt}$ which is the observed economic outcome for location $l$ in time period $t$.  We also define treated potential outcomes, $Y_{lt}(1)$, and untreated potential outcomes, $Y_{lt}(0)$, and note that $Y_{lt} = D_l Y_{lt}(1) + (1-D_l)Y_{lt}(0)$. The target parameter in this section is given by
\begin{align*}
    ATT^Y_t = \E[Y_t(1) - Y_t(0) | D=1]
\end{align*}
which is the difference between treated potential outcomes and untreated potential outcomes on average, in time period $t$, and among treated locations.  As discussed above, difference-in-differences is closely related to two-way fixed effects models for untreated potential outcomes, and, in this section, we consider the following model for untreated potential outcomes
\begin{align} \label{eqn:economic-outcomes-twfe}
    Y_{lt}(0) = \tau_t + \xi_l + \alpha I_{lt}(0) + v_{lt}
\end{align}
where $\tau_t$ is a common macro shock.  For economic outcomes, there is clear evidence of common macroeconomic shocks which can be motivated by, for example, common information about the health risks of Covid-19 across locations.  $\xi_l$ is a location-specific fixed effect allowing for time-invariant location-specific differences in economic outcomes, and $v_{lt}$ are idiosyncratic, time varying unobservables.

What is different about this model from standard DID is the term involving $I_{lt}(0)$ where $I_{lt}(0)$ is the number of Covid-19 cases in location $l$ in time period $t$ if the policy were not implemented.  It is likely to be very important to include this sort of term during the pandemic as it allows for economic outcomes to depend on the local spread of cases.  In particular, this allows for current cases to directly affect outcomes as well as individuals and/or firms taking more Covid-19 related precautions when the number of local cases is high.

In this section, we propose an approach that is able to deliver consistent estimates of $ATT^Y_t$ in the case when policies can have an effect on current Covid-19 cases and current Covid-19 cases can, in turn, have an effect on the outcome of interest.  Throughout this section, we contrast our suggested approach with two very common DID-type approaches.  First, we consider the case where a researcher compares the path of outcomes of treated locations to the path of outcomes among untreated locations without accounting for the current number of Covid-19 cases.  Throughout this section, we refer to this case as ``standard DID''.  Second, we consider a version of DID that includes the number of cases as a regressor.  Throughout this section, we refer to this case as ``regression DID''.  We show that both of these approaches generally deliver biased estimates of $ATT^Y_t$ under the model in \Cref{eqn:economic-outcomes-twfe}.  In the standard DID case, biased estimates arise because the researcher does not account for current cases in a particular location having a direct effect on outcomes.  In the regression DID case, biased estimates arise because the approach does not accommodate the possibility that the policy has an effect on the current number of cases (which in turn has an effect on outcomes).  

In light of this discussion, we propose an alternative approach that simultaneously addresses both of these issues.  We call our approach ``adjusted regression DID''.  Our idea is to include an adjustment term that accounts for the possibility that the policy affects the current number of cases.  This adjustment term is closely related to the arguments in the previous section; in particular, we can recover an estimate of the number of active cases that a treated location would experience in a particular time period by recovering the number of active cases in untreated locations with similar pre-treatment pandemic-related characteristics.  

Before stating the main result in this section, it is helpful to notice that, in the model in \Cref{eqn:economic-outcomes-twfe}, 
\begin{align} \label{eqn:economic-outcome-paths}
    \E[\Delta^{(t^*-1,t)}Y_t(0) | D=d] = \tilde{\tau}_t + \alpha \E[\Delta^{(t^*-1,t)} I_t(0) | D=d]
\end{align}
where we define $\tilde{\tau}_t := (\tau_t - \tau_{t^*-1})$.  Also note that $\tilde{\tau}_t$ and $\alpha$ are both identified using the untreated group (in that case, untreated potential outcomes and untreated potential active cases are observed in all time periods which implies that the parameters are identified as this amounts to a simple linear regression of $\Delta^{(t^*-1,t)} Y_t$ on $\Delta^{(t^*-1,t)} I_t$ using untreated locations).  

Next, to fix ideas, under standard DID, the estimator of $ATT^Y_t$ is the sample analogue of
\begin{align*}
    \E[\Delta^{(t^*-1,t)} Y_t | D=1] - \E[\Delta^{(t^*-1,t)} Y_t | D=0]
\end{align*}
Likewise, under regression DID, the estimator of $ATT^Y_t$ is the sample analogue of 
\begin{align*}
    \E[\Delta^{(t^*-1,t)} Y_t | D=1] - \big(\tilde{\tau}_t + \alpha \E[\Delta^{(t^*-1,t)} I_t | D=1] \big)
\end{align*}
Including covariates in this sort of way is a common strategy\footnote{Notice that this estimand is similar in spirit, though not exactly the same, as two way fixed effects regressions that include a treatment dummy variable along with other time varying covariates.  Besides the issues pointed out in this section (related to the covariates), those sorts of regressions do not generally deliver an interpretable treatment effect parameter in the case with multiple time periods and variation in treatment timing (see, for example, \citet{goodman-2021}).  The estimand mentioned above avoids the issues related to multiple periods and variation in treatment timing but, as we point in this section, still suffers from issues stemming from actual cases in treated locations not being equal to what cases would have been if the policy had not been implemented.} and would amount to comparing paths of outcomes for treated and untreated locations that experienced the same change in cases over time.  %

The next result provides an alternative identification result for $ATT^Y_t$ as well as results for the bias of standard DID and regression DID.
\begin{theorem} \label{thm:atty} In the model for untreated potential outcomes in  \Cref{eqn:economic-outcomes-twfe} and the  \Cref{ass:sird-model} and under \Cref{ass:overlap}, and for any $t \geq t^*$,  
    \begin{align} \label{eqn:path-current-cases}
        \E[\Delta^{(t^*-1,t)} I_t(0) | D=1] = \E\left[ \frac{D}{\E[D]} \Delta^{(t^*-1,t)} I_t - \omega(D,\mathcal{F}_{t^*-1}) (I_t - m^I_{0,t}(\mathcal{F}_{t^*-1})) \right]
    \end{align}
    where $m_{0,t}^I(\mathcal{F}_{t^*-1}) := \E[I_t | \mathcal{F}_{t^*-1}, D=0]$ and $\omega$ are the same weights as in \Cref{thm:unc}.  Moreover,
    \begin{align} \label{eqn:atty}
        ATT^Y_t = \E[\Delta^{(t^*-1,t)} Y_t | D=1] - \left\{ \tilde{\tau}_t + \alpha\left(\E[\Delta^{(t^*-1,t)} I_t(0) | D=1]\right) \right\}
    \end{align}
    and all the terms on the RHS of the expression for $ATT^Y_t$ are identified.  In addition, standard difference-in-differences is biased with bias given by
    \begin{align*}
        \E[\Delta^{(t^*-1,t)} Y_t | D=1] - \E[\Delta^{(t^*-1,t)} Y_t | D=0] - ATT^Y_t = \alpha \big(\E[\Delta^{(t^*-1,t)} I_t(0) | D=1] - \E[\Delta^{(t^*-1,t)} I_t(0) | D=0] \big)
    \end{align*}
    and the bias of regression DID (i.e., including current cases as a covariate) is given by
    \begin{align*}
        & \E[\Delta^{(t^*-1,t)} Y_t | D=1] - \big(\tilde{\tau}_t + \alpha \E[\Delta^{(t^*-1,t)} I_t | D=1]\big) - ATT^Y_t = - \alpha ATT^I_t
    \end{align*}
    where $ATT^I_t := \E[I_t(1) - I_t(0) | D=1]$.
\end{theorem}

It is worth sketching the arguments underlying the result in \Cref{thm:atty}.  To start with, notice that
\begin{align*}
    ATT^Y_t &= \E[\Delta^{(t^*-1,t)} Y_t | D=1] - \E[\Delta^{(t^*-1,t)}Y_t(0) | D=1]
\end{align*}
The bias of standard DID arises from (incorrectly) setting  $\E[\Delta^{(t^*-1,t)}Y_t(0) | D=1] = \E[\Delta^{(t^*-1,t)}Y_t | D=0]$.  In general, this sort of substitution is not appropriate because the path of outcomes that treated locations would have experienced in the absence of participating in the treatment depends on the path of active cases (which is not accounted for here).

Next, based on the model in \Cref{eqn:economic-outcomes-twfe}, it follows from \Cref{eqn:economic-outcome-paths} that
\begin{align} \label{eqn:untreated-economic-outcome-path}
    \E[\Delta^{(t^*-1,t)}Y_t(0) | D=1] = \tilde{\tau}_t + \alpha \E[\Delta^{(t^*-1,t)} I_t(0) | D=1]
\end{align}
The bias of regression DID (that directly includes current cases as a covariate) comes from (incorrectly) setting $\E[\Delta^{(t^*-1,t)} I_t(0) | D=1] = \E[\Delta^{(t^*-1,t)} I_t | D=1]$.  This strategy is also not generally appropriate because the policy can change (and is likely targeted at changing) the path of active cases.

By contrast, our approach uses the expression in \Cref{eqn:path-current-cases} for $\E[\Delta^{(t^*-1,t)} I_t(0)|D=1]$.  This expression takes the observed path of active cases and subtracts from it the effect of the policy on active cases (which is the term $\E[\omega(D,\mathcal{F}_{t^*-1}) (I_t - m^I_{0,t}(\mathcal{F}_{t^*-1}))]$ and holds under the \Cref{ass:sird-model}).  Notice that this term is analogous to the expression for the effect of the policy on cumulative cases in \Cref{thm:unc} in the previous section.  The difference between the observed path of active cases among treated locations and the effect of the policy on active cases recovers the path of active cases that would have occurred if the policy had not been implemented.  Given this expression, it can be plugged into \Cref{eqn:untreated-economic-outcome-path} to recover the path of untreated potential outcomes and, hence, to recover $ATT^Y_t$.

\paragraph{Estimation:} The above discussion suggests the following estimation strategy:
\begin{enumerate}
    \item Run a regression of $\Delta^{(t^*-1,t)} Y_{lt}$ on $\Delta^{(t^*-1,t)} I_{lt}$ using untreated locations in order to estimate $\tilde{\tau}_t$ and $\alpha$.  
    \item Estimate the propensity score, $p(\mathcal{F}_{t^*-1})$, using the entire sample.  Plug these estimates into the sample analogue of \Cref{eqn:weights} in order to estimate the weights $\omega$.  Estimate $m_{0,t}^I(\mathcal{F}_{t^*-1})$ using untreated locations.  Plug in the estimates of $\omega$ and $m_{0,t}^I$ into the sample analogue of \Cref{eqn:path-current-cases} to compute an estimate of $\E[\Delta^{(t^*-1,t)} I_t(0) | D=1]$.
    \item Plug in the preliminary estimators in Steps 1 and 2 to estimate $ATT^Y_t$ directly using the expression in \Cref{thm:atty}.
\end{enumerate}

In the Supplementary Appendix, we provide the asymptotic distribution of our estimator of $ATT^Y_t$.  The estimation procedure involves several steps, but each step is parametric and the limiting distribution of the estimate of $ATT^Y_t$ can be obtained following well-known arguments about multiple step estimation procedures that account for estimation effects of each step.  In particular, the term $\E[\omega(D,\mathcal{F}_{t^*-1}) (I_t - m_{0,t}^I(\mathcal{F}_{t^*-1}))]$ can be handled using exactly the same arguments as in the previous section.  The other steps in the estimation procedure only involve either running simple parametric regressions or directly calculating averages and are therefore straightforward to account for.  As earlier, in practice, we use the multiplier bootstrap to conduct inference and discuss how to conduct uniform inference across different time periods.

\begin{remark}
In general, it is not possible to sign the bias from using standard DID or regression DID (as an extreme example, over long enough time horizons, a policy that is effective at slowing the spread of Covid-19 could lead to higher current infections if untreated locations reach herd immunity).  That said, over relatively short time horizons, it is possible to get a sense of the \textit{likely} directions of bias.  In particular, suppose that (i) $\alpha < 0$ so that more current cases lead to lower economic outcomes, (ii) that the time horizon is short and the policy decreases the number of active cases over a short horizon, (iii) that the pandemic is in its early stages and that treated locations tend to have earlier arrival times of their first cases (so that, in the absence of participating in the treatment, treated locations would have experienced larger increases in the number of active cases than untreated locations), and (iv) the policy has a negative effect on economic outcomes.  In this case, both standard DID and regression DID (that adjusts for the actual number of cases) will both overstate the magnitude of the effect of the policy.
\end{remark}

\section{Monte Carlo Simulations}

In this section, we provide some Monte Carlo simulations to demonstrate the performance of the main estimation strategies considered in the paper.  To begin with, we consider estimating the effect of a policy on cumulative Covid-19 cases.  In order to generate the data, we consider the case where untreated potential outcomes are generated by the \ref{ass:sird-model}.  The values for the main parameters in the SIRD model are provided in \Cref{tab:sim-params} in \Cref{app:sim-details}.  We also suppose that the policy has no effect on the pandemic so that all treatment effects are equal to 0.

Throughout this section, we consider the case where there are 250 locations (we vary this number in a few cases), where the probability of a location being treated is equal to 0.5, and where there are 1000 individuals in each location.  We report bias, root mean squared error, and rejection probabilities for $H_0: ATT=0$ for the average effect of the policy across the first 50 post-treatment time periods (i.e., we compute event-study type estimates for 50 periods following the treatment, average them across event time to get an overall average treatment effect parameter, and compute the properties of this estimator).  To implement our doubly robust estimator, we include a third order polynomial (also including all interactions) in the pre-treatment number of infected individuals and pre-treatment number of susceptible individuals both for the outcome regression and for the propensity score.  Across simulations, we primarily focus on varying the timing of initial Covid-19 cases among treated and untreated locations, and on varying the treatment timing across treated and untreated locations.

{ \setlength{\tabcolsep}{10pt}
\begin{table}[t]
\centering
\caption{Monte Carlo Simulations for Cumulative Cases}
\label{tab:mc_cum_cases}
\resizebox{\linewidth}{!}{
\begin{tabular}[t]{rrrrrrrrr}
\toprule
\multicolumn{3}{c}{ } & \multicolumn{3}{c}{Unconfoundedness} & \multicolumn{3}{c}{DID} \\
\cmidrule(l{3pt}r{3pt}){4-6} \cmidrule(l{3pt}r{3pt}){7-9}
Policy Time & $\lambda_D$ & $\lambda_U$ & Bias & RMSE & Rej.\,Prob. & Bias & RMSE & Rej.\,Prob.\\
\midrule
\addlinespace[0.3em]
\multicolumn{9}{l}{\textbf{Vary Treated First Case Timing}}\\
\hspace{1em}\cellcolor{white}{150} & \cellcolor{white}{40} & \cellcolor{white}{60} & \cellcolor{white}{0.009} & \cellcolor{white}{0.478} & \cellcolor{white}{0.039} & \cellcolor{white}{-3.044} & \cellcolor{white}{4.169} & \cellcolor{white}{0.162}\\
\hspace{1em}150 & 60 & 60 & 0.008 & 0.582 & 0.044 & 0.031 & 3.233 & 0.055\\
\hspace{1em}\cellcolor{white}{150} & \cellcolor{white}{80} & \cellcolor{white}{60} & \cellcolor{white}{-0.012} & \cellcolor{white}{0.750} & \cellcolor{white}{0.065} & \cellcolor{white}{2.931} & \cellcolor{white}{4.542} & \cellcolor{white}{0.153}\\
\addlinespace[0.3em]
\multicolumn{9}{l}{\textbf{Vary Policy Timing}}\\
\hspace{1em}75 & 40 & 80 & 0.034 & 0.803& 0.036 & -12.829 & 14.416 & 0.469\\
\hspace{1em}\cellcolor{white}{150} & \cellcolor{white}{40} & \cellcolor{white}{80} & \cellcolor{white}{0.034} & \cellcolor{white}{0.428} & \cellcolor{white}{0.024} & \cellcolor{white}{-5.593} & \cellcolor{white}{6.464} & \cellcolor{white}{0.438}\\
\hspace{1em}225 & 40 & 80 & 0.047 & 0.196 & 0.031 & -1.133 & 1.389 & 0.323\\
\addlinespace[0.3em]
\multicolumn{9}{l}{\textbf{Vary Number of Locations, $n=1000$}}\\
\hspace{1em}\cellcolor{white}{150} & \cellcolor{white}{40} & \cellcolor{white}{80} & \cellcolor{white}{0.031} & \cellcolor{white}{0.194} & \cellcolor{white}{0.044} & \cellcolor{white}{-5.680} & \cellcolor{white}{5.895} & \cellcolor{white}{0.951}\\
\bottomrule
\end{tabular}}
\begin{justify}
{ \footnotesize \textit{Notes:}  The table provides Monte Carlo simulations for cumulative cases using the unconfoundedness approach suggested in the paper as well as difference-in-differences and with the simulation parameters discussed in the text.  The column labeled ``Policy Time'' indicates the timing when the policy is implemented among treated locations; $\lambda_D$ and $\lambda_U$ are the mean timing of the first case for treated and untreated locations, respectively.  The other columns report the bias, root mean squared error (RMSE), and rejection probabilities for each simulation setup.  }
\end{justify}
\end{table}
}

The results for our first set of simulations are provided in \Cref{tab:mc_cum_cases}.   The high level takeaway from this table is that the unconfoundedness approach uniformly appears to perform better than difference-in-differences.  Difference-in-differences is severely biased when the timing of initial cases is different between treated and untreated locations (this is in line with our earlier discussion).  The magnitude of the bias of difference-in-differences is also sensitive to the timing of the policy (this holds because the direction/magnitude of violations of parallel trends depends on the shape of pandemic related variables which are, in turn, dependent on how long ago the pandemic started).  Across simulations, difference-in-differences also tends to over-reject.

On the other hand, the doubly robust unconfoundedness approach performs much better with good performance across each specification.  Interestingly, even in the case where the first cases show up, on average, at the same time across treated and untreated locations (in this case, as expected, DID appears to be unbiased), the unconfoundedness approach suggested in the paper has notably smaller root mean squared error.

{ \setlength{\tabcolsep}{10pt}
\begin{table}[t]
\centering
\caption{Monte Carlo Simulations for Economic Outcomes}
\label{tab:mc_economic_outcome}
\begin{tabular}[t]{rrrrrrrrr}
\toprule
\multicolumn{3}{c}{ } & \multicolumn{3}{c}{Adjusted Regression DID} & \multicolumn{3}{c}{Standard DID} \\
\cmidrule(l{3pt}r{3pt}){4-6} \cmidrule(l{3pt}r{3pt}){7-9}
Policy Time & $\lambda_D$ & $\lambda_U$ & Bias & RMSE & Rej.\,Prob. & Bias & RMSE & Rej.\,Prob.\\
\midrule
\addlinespace[0.3em]
\multicolumn{9}{l}{\textbf{$n=250$}}\\
\hspace{1em}\cellcolor{white}{150} & \cellcolor{white}{40} & \cellcolor{white}{60} & \cellcolor{white}{0.000} & \cellcolor{white}{0.127} & \cellcolor{white}{0.048} & \cellcolor{white}{-0.134} & \cellcolor{white}{0.227} & \cellcolor{white}{0.092}\\
\hspace{1em}150 & 60 & 60 & 0.001 & 0.132 & 0.048 & 0.002 & 0.193 & 0.043\\
\hspace{1em}\cellcolor{white}{150} & \cellcolor{white}{80} & \cellcolor{white}{60} & \cellcolor{white}{-0.015} & \cellcolor{white}{0.132} & \cellcolor{white}{0.049} & \cellcolor{white}{0.110} & \cellcolor{white}{0.230} & \cellcolor{white}{0.081}\\
\addlinespace[0.3em]
\multicolumn{9}{l}{\textbf{$n=1000$}}\\
\hspace{1em}150 & 40 & 60 & 0.003 & 0.066 & 0.055 & -0.129 & 0.159 & 0.263\\
\hspace{1em}\cellcolor{white}{150} & \cellcolor{white}{60} & \cellcolor{white}{60} & \cellcolor{white}{0.005} & \cellcolor{white}{0.067} & \cellcolor{white}{0.045} & \cellcolor{white}{0.005} & \cellcolor{white}{0.098} & \cellcolor{white}{0.051}\\
\hspace{1em}150 & 80 & 60 & 0.001 & 0.071 & 0.068 & 0.127 & 0.165 & 0.240\\
\bottomrule
\end{tabular}
\begin{justify}
{ \footnotesize \textit{Notes:}  The table provides Monte Carlo simulations for economic outcomes using the adjusted regression DID approach suggested in the paper as well as standard DID and with the simulation parameters discussed in the text.  The column labeled ``Policy Time'' indicates the timing when the policy is implemented among treated locations; $\lambda_D$ and $\lambda_U$ are the mean timing of the first case for treated and untreated locations, respectively.  The other columns report the bias, root mean squared error (RMSE), and rejection probabilities for each simulation setup.  }
\end{justify}
\end{table}
}

Next, we provide analogous results but for the effect of the policy on economic outcomes.  For this part, we generate untreated potential outcomes according to \Cref{eqn:economic-outcomes-twfe}.  We set $\alpha=-0.1$, $\tau_t = (50 + 20 \times t/ \mathcal{T})$, and we set $\eta|D=d \sim N(\mu_d, 1)$ where $\mu_d = 20 - 10d$ for $d \in \{0,1\}$.  We also set the parameters for the pandemic related variables as in the baseline specification discussed above. 

These results are provide in  \Cref{tab:mc_economic_outcome} where we vary the timing of initial cases as well as the number of locations across simulations.  As in the previous case, the approach suggested in the paper (adjusted regression DID) performs well uniformly across DGPs.  By contrast, standard DID performs less well particularly in cases where the timing of initial cases is systematically different across treated and untreated locations.

\section{Application: Effects of Shelter-in-Place Orders on Covid-19 Cases and Travel} \label{sec:application}

To conclude the paper, we apply our approach to study the effect of shelter-in-place orders (SIPOs) on Covid-19 cases and travel.  We start by using state-level data to evaluate the effect of SIPOs on Covid-19 cases.  This approach is broadly similar to a number of other papers including   \citet{berry-fowler-glazer-handel-macmillen-2021}, \citet{bendavid-oh-bhattacharya-ioannidis-2021}, \citet{courtemanche-garuccio-le-pinkston-yelowitz-2020}, \citet{dave-friedson-matsuzawa-sabia-2021}, \citet{dave-friedson-matsuzawa-sabia-safford-2020}, \citet{hsiang-et-al-2020}, and \citet{villas-sears-villas-villas-2020}, among others.  We consider a number of variations of difference-in-differences estimation strategies in this context as well as implementing the  unconfoundedness approach discussed above.  Importantly, we document substantial differences in important pre-treatment characteristics such as the number of Covid-19 cases between states that were early adopters of SIPOs, late adopters of SIPOs, or never implemented a SIPO.  As emphasized above, this suggests that the parallel trends assumption underlying the DID approach is likely to be violated.  We find that DID approaches can, in some cases, lead to unreasonable estimates that SIPOs \textit{increased} Covid-19 cases.  Another of our main findings is that the DID approach is quite sensitive to seemingly minor modifications to the specification such as using the logarithm or level of the outcome or whether or not one includes location-specific linear trends.  

For any approach using state-level data, there are some important challenges.  One of these is that a number of additional Covid-19 related policies such as emergency declarations, school closures, and business closures were often implemented around the same time.  Moreover, there is variation across states both in terms of exactly which policies were implemented as well as the timing of these policies; we discuss some other challenges below as well.  Therefore, as a second step, we use county-level data and provide difference-in-differences and unconfoundedness estimates of policy effects in states where a SIPO was implemented relative to bordering states that never implemented a SIPO among states that are similar both in terms of their pre-treatment pandemic-related characteristics (such as having experienced a similar number of Covid-19 cases) and in terms of the mix and timing of other policies that were implemented.

In general, using variations of DID, we tend to find a hard-to-interpret mix of policy effects that include estimates indicating large reductions in Covid-19 cases due to SIPOs, no effect of SIPOs, or even relatively large \textit{increases} in Covid-19 cases due to SIPOs.  This contrasts with our results using unconfoundedness which are more consistent across different state-level policies and where we tend to find relatively smaller reductions in Covid-19 cases due to SIPOs.  Finally, we also use the county-level data to study the effects of SIPOs on travel.  In line with the literature (e.g., \citet{goolsbee-syverson-2021}), we tend to find relatively small reductions in travel due to SIPOs.  

\subsection{State-Level Results}

Our first set of results come from analyzing state-level data.  We consider a period early in the pandemic --- March 10, 2020 to May 1, 2020 --- when a large number of states implemented shelter-in-place orders.  

\subsubsection*{Data} 

We follow \citet{dave-friedson-matsuzawa-sabia-2021} in terms of definitions of shelter-in-place orders and the timing of implementation across states.  In order to facilitate estimating conditional treatment assignment probabilities, we assign states into ``groups'' on the basis of the timing when they adopted a SIPO.  And, in particular, we assign states that adopted a SIPO within a five day window, starting on March 19, into the same group.  For example, California was the first state to implement a shelter-in-place order on March 19; Illinois and New Jersey followed on March 21; New York on March 22; and Connecticut, Louisiana, Oregon, and Washington on March 23.  These form a group of states that we refer to as the March 19 group.   Sixteen other states adopted shelter-in-place orders between March 24 and March 28 and form the group that we refer to as the March 24 group.  We include four such groups total as well as an untreated group of ten states that did not adopt a shelter-in-place order over the time period that we consider.

Next, we obtained data on state-level Covid-19 cases and testing from the Centers for Disease Control COVID Data Tracker (\url{https://covid.cdc.gov/covid-data-tracker/}).  We also use 2019 state-level populations from the Census Bureau.  In order to deal with heterogeneity in terms of state populations, we use versions of pandemic related variables in terms of their number per thousand individuals in a particular state (e.g., cumulative cases per thousand individuals).  In terms of the SIRD model, this amounts to dividing all variables by $N_l$ and multiplying by one thousand; this transformation is compatible with the SIRD model.  We construct the current number of active Covid-19 cases (and therefore contagious individuals) as the total number of newly reported cases over the past seven days; as for the other pandemic related variables, we use the number of current cases per thousand individuals in a state.   Finally, we use travel data from Google's Covid-19 Community Mobility Report (\url{https://www.google.com/covid19/mobility}).  We focus on state-level retail and recreation travel (these are aggregated together) which is reported as a percentage change relative to pre-Covid travel baselines.

{ \setlength{\tabcolsep}{16pt}
\begin{table}[t]
\centering
\caption{Summary Statistics for State-Level Data}
\label{tab:ss}
\begin{tabular}{lrrrrr}
\toprule
\multicolumn{1}{c}{} & \multicolumn{5}{c}{Group} \\
\cmidrule(l{3pt}r{3pt}){2-6}
 & Untreated & Mar 19 & Mar 24 & Mar 29 & Apr 3 \\
\midrule
group size & 10 & 8 & 16 & 10 & 6 \\
pop.\,(millions) & 2.9 & 12.6 & 3.7 & 8.8 & 8.5 \\
\addlinespace[0.8em]
\multicolumn{6}{l}{\underline{Cases:}}\\[5pt]
\hspace{1em}Mar 14 & 12.2 & 15.0 & 15.3 & 3.1 & 17.9\\
\hspace{1em}Mar 21 & 52.3 & 114.6 & 82.4 & 27.9 & 89.7\\
\hspace{1em}Mar 28 & 190.0 & 573.3 & 257.3 & 144.0 & 267.8\\
\hspace{1em}Apr 4 & 440.8 & 1609.6 & 566.0 & 392.3 & 555.6\\
\hspace{1em}Apr 11 & 807.0 & 2742.6 & 944.1 & 739.7 & 902.0\\
\hspace{1em}Apr 18 & 1285.1 & 3696.1 & 1324.1 & 1075.2 & 1223.4\\
\hspace{1em}Apr 25 & 1917.6 & 4701.9 & 1765.3 & 1475.7 & 1567.1\\
\addlinespace[0.8em]
\multicolumn{6}{l}{\underline{Travel}}\\
\hspace{1em}Mar 14 & -9.8 & -10.5 & -6.9 & -5.7 & -1.3\\
\hspace{1em}Mar 21 & -39.4 & -43.1 & -40.4 & -38.7 & -34.0\\
\hspace{1em}Mar 28 & -44.5 & -53.5 & -52.9 & -45.2 & -39.8\\
\hspace{1em}Apr 4 & -45.7 & -52.6 & -49.0 & -47.9 & -46.5\\
\hspace{1em}Apr 11 & -42.2 & -49.8 & -45.8 & -42.7 & -41.3\\
\hspace{1em}Apr 18 & -36.9 & -52.0 & -43.8 & -41.9 & -36.3\\
\hspace{1em}Apr 25 & -33.2 & -47.5 & -38.5 & -39.2 & -34.2\\
\bottomrule
\end{tabular}

\begin{justify}
\footnotesize{\textit{Notes:} The table provides summary statistics for the state-level data used in the application.  A state's ``group'' is defined by the time period when it became treated rounded to the nearest 5th day (see text for a detailed explanation).  Cases are reported as the number per million individuals in a state and averaged within the corresponding ``group.''  Travel data is reported as the percentage change in retail and recreation trips relative to pre-Covid baseline.}
\end{justify}
\end{table}
}

Summary statistics for the data that we use are provided in \Cref{tab:ss}.  There are some things that are immediately notable from the summary statistics.  First, early in the pandemic, the number of cases were substantially different for states that adopted shelter-in-place orders earlier relative to states that adopted them later or that did not adopt them at all.  This immediately suggests that it will be challenging for DID to perform well at evaluating the effect of shelter-in-place orders on the number of cases.  In addition, notice that early treated states (particularly, the March 19 group) experienced very large increases in their number of cases relative to later- and never-adopters of the policy.  %
Finally, the second panel of the table shows changes in retail and recreation trips.  The most notable feature of this part of the table is that there were large decreases in travel across all states regardless of their shelter-in-place policies.

\subsubsection*{Results}

\begin{figure}[t!]
    \centering
    \caption{Policy Effects of SIPOs on Covid-19 Cases using TWFE Event Study Regressions}
    \label{fig:twfe-state-results}
    \begin{subfigure}[b]{.49\textwidth}
    \includegraphics[width=\textwidth]{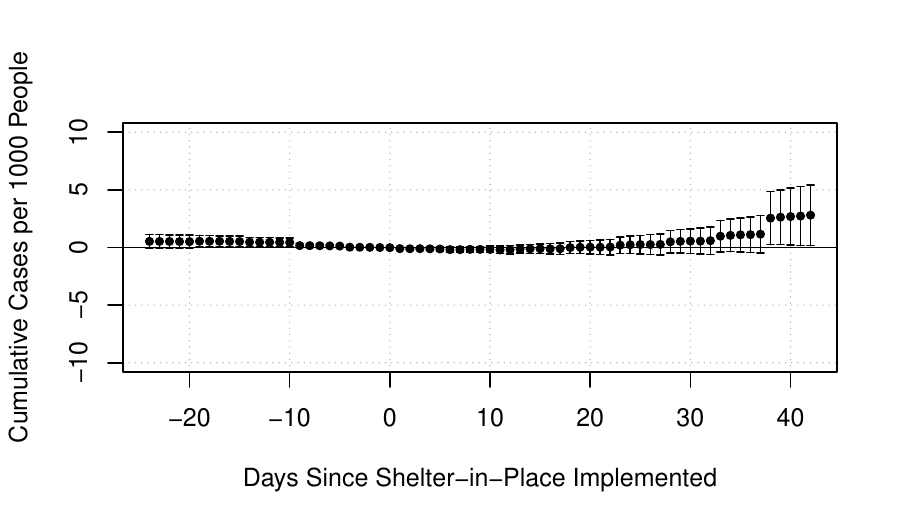}
    \caption{Outcome: Level, Linear Trend: No}
    \end{subfigure}
    \begin{subfigure}[b]{.49\textwidth}
    \includegraphics[width=\textwidth]{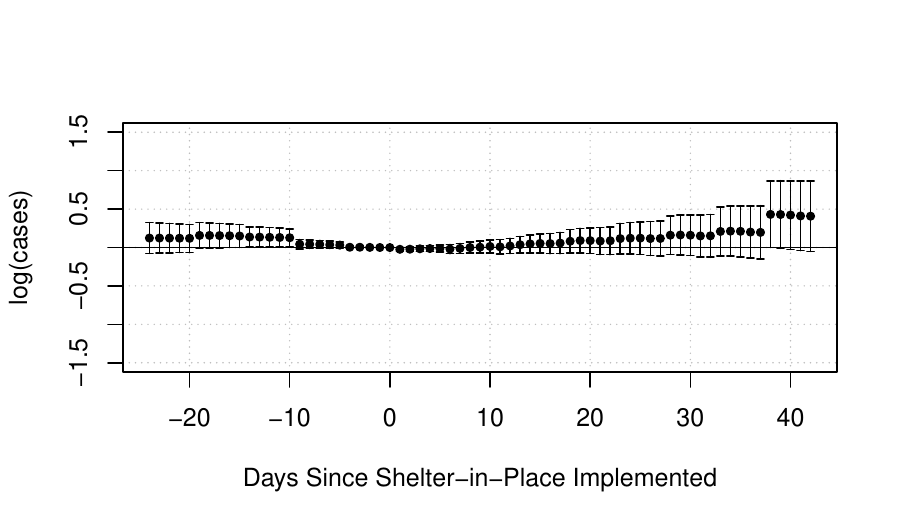}
    \caption{Outcome: Log, Linear Trend: No}
    \end{subfigure}
    \begin{subfigure}[b]{.49\textwidth}
    \includegraphics[width=\textwidth]{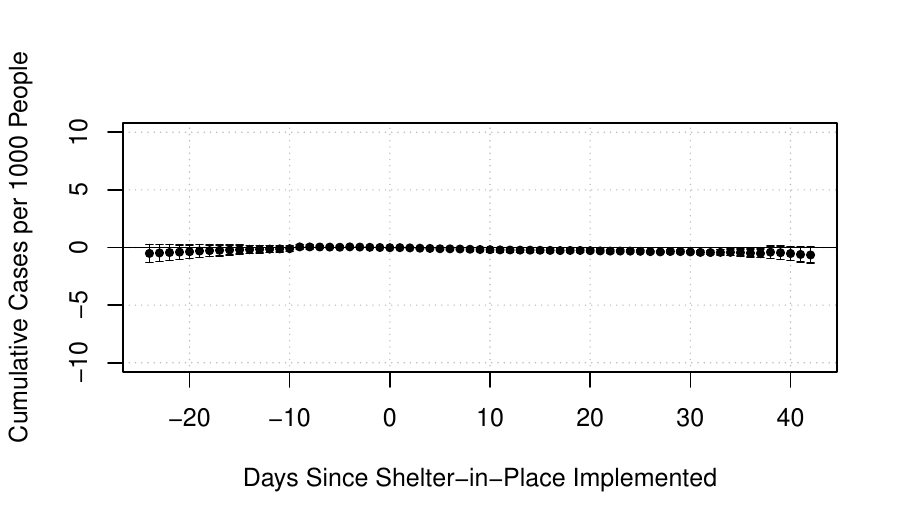}
    \caption{Outcome: Level, Linear Trend: Yes}
    \end{subfigure}
    \begin{subfigure}[b]{.49\textwidth}
    \includegraphics[width=\textwidth]{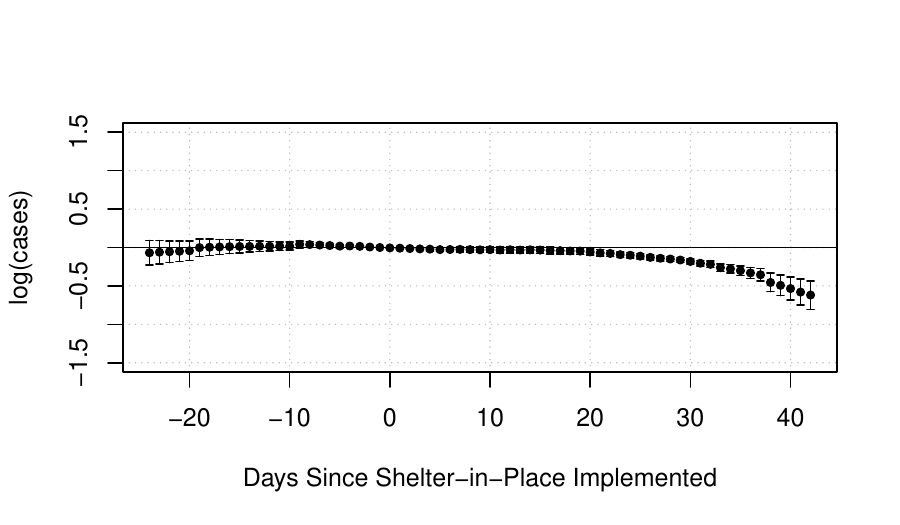}
    \caption{Outcome: Log, Linear Trend: Yes}
    \end{subfigure}
    
    \begin{justify}
    { \footnotesize \textit{Notes:}  The figure contains event study type estimates of the effect of SIPOs on the number of cumulative Covid-19 cases that come from a two-way fixed effects regression.  $e=0$ corresponds to the time period when the policy was implemented.  Negative values of $e$ correspond to pre-treatment estimates of the effect of the policy (and can be thought of as pre-tests), and positive values of $e$ correspond to estimates of the effect of the policy at different lengths of exposure to the treatment.  The estimates are normalized to be equal to 0 for $e=-1$.  The estimates across panels differ based on (i) whether the outcome is in levels or in logarithms and (ii) whether or not estimates include location-specific linear trends.  90\% confidence intervals are provided by the vertical bars in each panel. }
    \end{justify}
\end{figure}

Our first set of results come from using state-level data and several different types of difference-in-differences estimation strategies.  In \Cref{fig:twfe-state-results}, we use two-way fixed effects (TWFE) event study regressions of the form
\begin{align} \label{eqn:twfe-es}
    C_{lt} = \theta_t + \eta_i + \sum_{e=-({\mathcal{T}-1)}}^{-2} \beta_e D_{lt}^e + \sum_{e=0}^{\mathcal{T}-1} \beta_e D_{lt}^e + v_{it}
\end{align}
where $D_{lt}^e$ is a binary variable that is equal to 1 for location $l$ in time period $t$ if that location has been treated for exactly $e$ periods in period $t$ and is otherwise equal to 0.  These sorts of regressions have been widely used in work evaluating the effects of Covid-19 related policies. The panels in \Cref{fig:twfe-state-results} differ along two dimensions: first, whether the outcome is the logarithm or the level of the number of Covid-19 cases; and second, whether or not the specification additionally includes a location-specific linear time trend.  All of these specifications are common in the literature, and, in particular, the specification where the outcomes is in logarithms and includes a location-specific linear trend is a main specification in \citet{dave-friedson-matsuzawa-sabia-2021}.

\Cref{fig:twfe-state-results} highlights that the qualitative conclusion as to whether or not  SIPOs affected the number of Covid-19 cases is highly sensitive to functional form assumptions made by the researcher.  First, when the outcome is the logarithm of the number of Covid-19 cases and the specification includes a location-specific linear time trend, then the estimates indicate a large reduction in Covid-19 cases due to shelter-in-place policies (see panel (d) of \Cref{fig:twfe-state-results}).  The results in panel (c), which include also include a linear trend but where the outcome is the level of the number of cases, also indicate that SIPOs may have reduced the number of Covid-19 cases (these results are closer to zero and only marginally statistically significant).    On the other hand, the results in panels (a) and (b) of \Cref{fig:twfe-state-results}, neither of which include a location-specific linear trend, are much different and suggest that SIPOs led to an \textit{increase} in the number of Covid-19 cases.  It seems very hard to rationalize these sorts of results; in particular, it would seem that shelter-in-place orders could either have no effect or decrease Covid-19 cases, but it is difficult to see how they could increase cases.  However, even from the summary statistics, one can see that DID estimates are likely to be positive as early policy adopters were tending to experience larger increases in cases.  In our view, a better explanation for these results is that Covid-19 was more prevalent earlier in locations that were early policy adopters and that the strong, early exponential growth of Covid-19 cases overwhelms any reduction in the infection rate due to the policy.  It is exactly in this case where DID would be susceptible to attributing faster growth in Covid-19 cases to the policy rather than to simply a larger number of early cases in treated locations. 

Another noteworthy feature of \Cref{fig:twfe-state-results} is that none of the four specifications used in the figure result in either large or statistically significant violations of the underlying parallel trends assumption in pre-treatment periods (i.e., estimates different from 0 for $e<0$).\footnote{To be more specific, there are 23 pre-treatment estimates reported in each panel.  No pre-treatment estimates are statistically significant at the 5\% level in panels (a), (b), or (c).  One estimate is statistically significant at the 5\% level in panel (d); this occurs for $e=-3$ and the p-value is 0.037.}  This suggests that it is not possible to use a purely data-driven/reduced form model selection procedure to infer that one specification is likely to perform better than the others, despite the choice of the model largely driving the results.\footnote{In one sense, it is somewhat surprising that that parallel trends cannot be rejected in pre-treatment periods for any of the estimation strategies considered here.  The \ref{ass:sird-model} implies that, for all of the models considered here, the parallel trends assumption is also violated in pre-treatment periods.  Instead, the results here indicate that we are not able to \textit{detect} violations of parallel trends in pre-treatment periods (even though parallel trends is probably actually violated).  That we are not able to detect violations of parallel trends is not altogether surprising as pre-tests are often under-powered (\citet{roth-2022}), and this is likely to be an acute issue early in the pandemic when the number of Covid-19 cases is very small in most pre-treatment periods.}  Finally, we should emphasize that \textit{none} of the TWFE event study specifications discussed here are compatible with the SIRD model that we have discussed in the paper.  

\begin{figure}[t!]
    \centering
    \caption{Policy Effects of SIPOs on Covid-19 Cases using Treatment Effect Heterogeneity-Robust Estimation Strategies}
    \label{fig:did-state-results}
    Panel (i): \citet{callaway-santanna-2021} \\
    \begin{subfigure}[b]{.49\textwidth}
    \includegraphics[width=\textwidth]{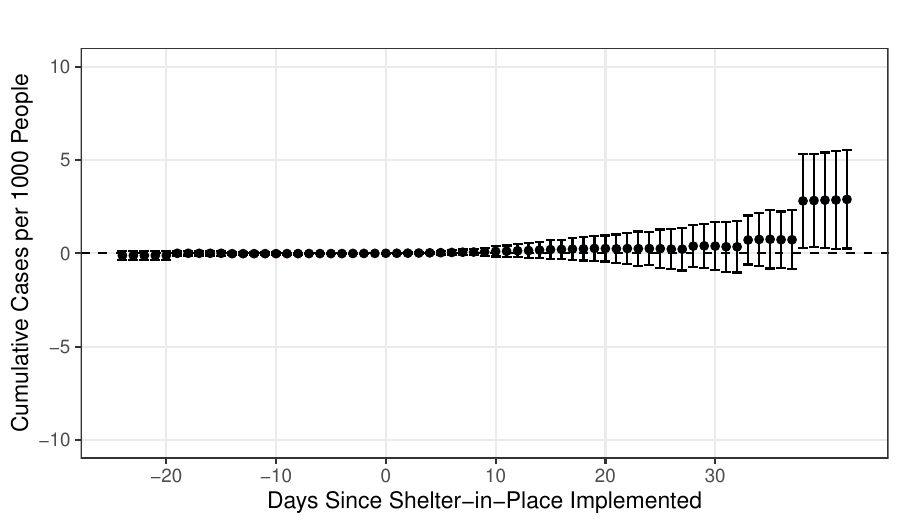}
    \caption{Outcome: Level, Linear Trend: No}
    \end{subfigure}
    \begin{subfigure}[b]{.49\textwidth}
    \includegraphics[width=\textwidth]{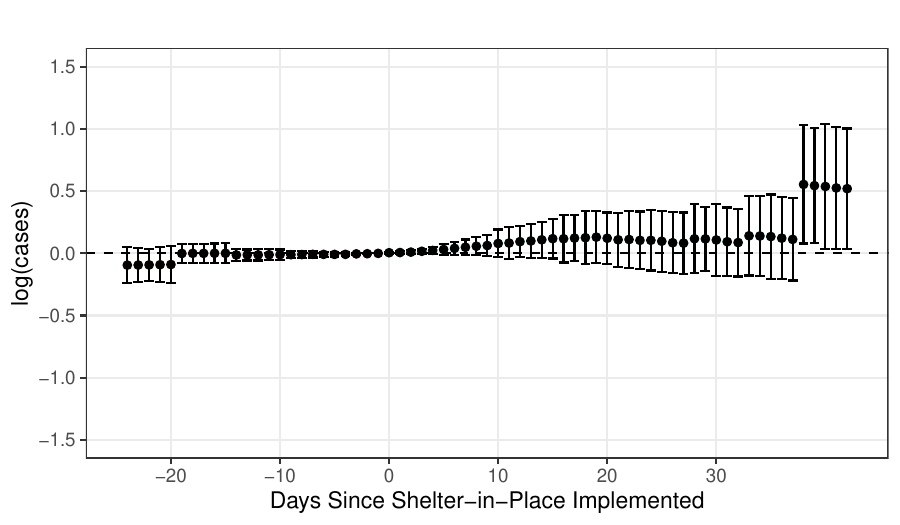}
    \caption{Outcome: Log, Linear Trend: No}
    \end{subfigure} \\[25pt]
    Panel (ii): \citet{gardner-2021} \\ 
    \begin{subfigure}[b]{.49\textwidth}
    \includegraphics[width=\textwidth]{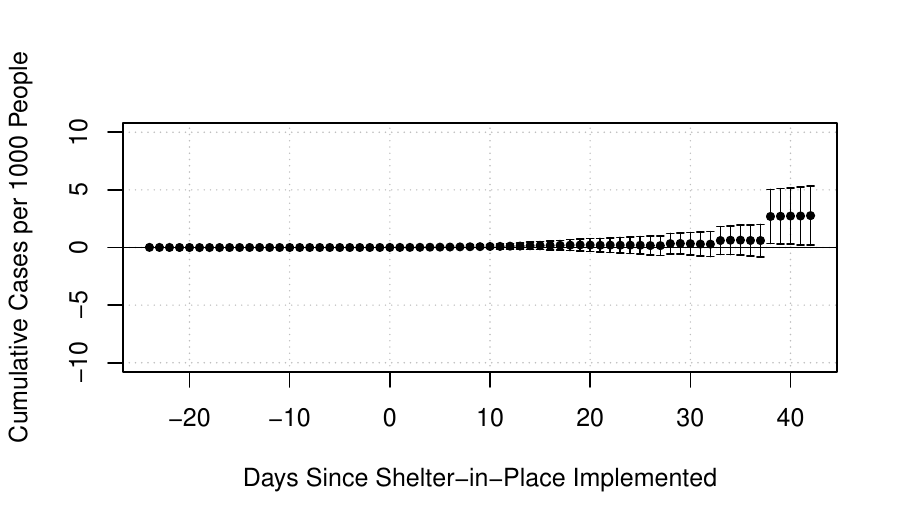}
    \caption{Outcome: Level, Linear Trend: Yes}
    \end{subfigure}
    \begin{subfigure}[b]{.49\textwidth}
    \includegraphics[width=\textwidth]{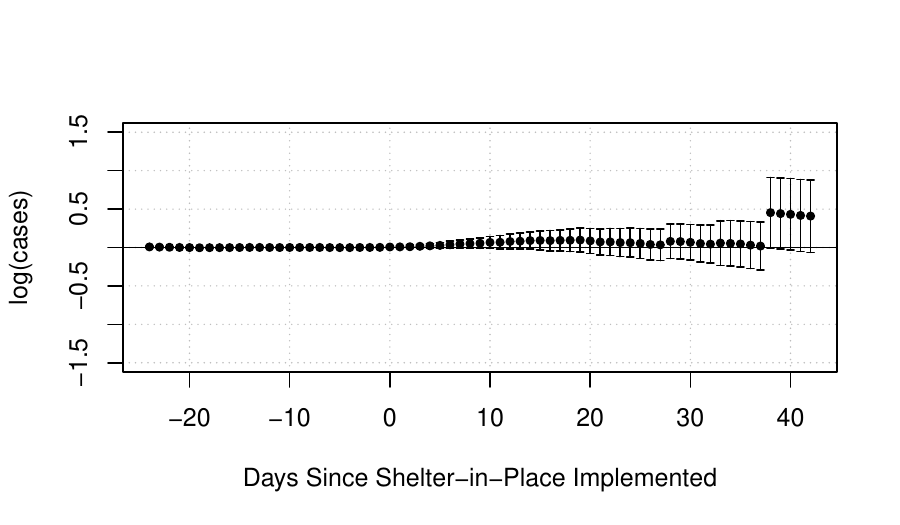}
    \caption{Outcome: Log, Linear Trend: Yes}
    \end{subfigure}
    
    \begin{justify}
    { \footnotesize \textit{Notes:}  The figure contains event study type estimates of the effect of SIPOs on the number of cumulative Covid-19 cases that come from several variations of heterogeneity-robust DID identification strategies (as discussed in the text).  $e=0$ corresponds to the time period when the policy was implemented.  Negative values of $e$ correspond to pre-treatment estimates of the effect of the policy (and can be thought of as pre-tests), and positive values of $e$ correspond to estimates of the effect of the policy at different lengths of exposure to the treatment.  The estimates across panels differ based on (i) whether the outcome is in levels or in logarithms and (ii) whether or not estimates include location-specific linear trends.  In Panels (a) and (b) the pre-treatment estimates have the same interpretation as the TWFE estimates in \Cref{fig:twfe-state-results}; on the other hand, the pre-treatment estimates in Panels (c) and (d), have a placebo interpretation (i.e., they provide the estimate of the ``effect of the treatment as if it had been implemented in that particular period'').  Thus, the confidence intervals are systematically narrower in these panels (see \citet{brown-butts-2022} for additional details along these lines). 90\% confidence intervals are provided by the vertical bars in each panel. }
    \end{justify}
\end{figure}

One possible concern with the previous results is related to limitations of two-way fixed effects regressions when there is variation in treatment timing and treatment effect heterogeneity (see, for example, \citet{chaisemartin-dhaultfoeuille-2020,goodman-2021,sun-abraham-2021,borusyak-jaravel-spiess-2021}).  There is certainly variation in treatment timing for SIPOs and heterogeneous effects (which would include things like variation in policy effects according the time period when the policy is adopted or effects that vary with length of exposure to the policy, among others) are very likely as well. These issues are considered in \citet{dave-friedson-matsuzawa-sabia-2021} and discussed extensively in \citet{goodman-marcus-2020}.  In order to address these possible concerns, \Cref{fig:did-state-results} provides estimates using ``heterogeneity robust'' DID estimation strategies from \citet{callaway-santanna-2021,gardner-2021}.  The setup for these results mirrors the setup from \Cref{fig:twfe-state-results} as the panels differ based on whether the outcome is in levels or logarithms and by whether or not the results include location-specific linear trends.  The results in panels (a) and (b) that do not include a linear time trend are broadly similar to the TWFE event study regressions from above but still suggest that SIPOs \textit{increased} Covid-19 cases.  On the other hand, the results that include location-specific linear trends are notably different from the results in \Cref{fig:twfe-state-results}.  Using heterogeneity robust versions of DID estimation strategies changes the sign of the estimates and results in estimates that SIPOs increased Covid-19 cases.  This is disconcerting as these estimation strategies provide a number of advantages relative to the TWFE event studies; however, as discussed above, it seems hard to understand how SIPOs could lead to an increase in Covid-19 cases.  As before, we do not find large or any statistically significant estimates of Covid-19 policies in pre-treatment periods.\footnote{Depending on the specification, there are either 23 or 24 pre-treatment estimates in each panel of \Cref{fig:did-state-results}.  Across all four panels, none of these estimates are statistically different from 0 at the 5\% level.}

\begin{figure}[t!]
    \centering
    \caption{Policy Effects of SIPOs on Covid-19 Cases under Unconfoundedness}
    \label{fig:state-covid-results-unc}
    \includegraphics[width=.6\textwidth]{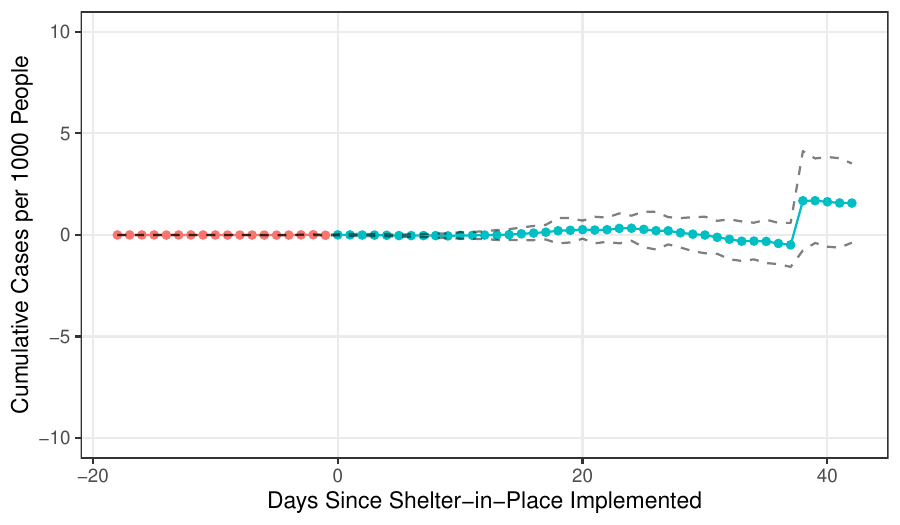}
    \begin{justify}
    { \footnotesize \textit{Notes:}  The figure contains event study type estimates of the effect of SIPOs on the number of cumulative Covid-19 cases uses the unconfoundedness approach discussed in the paper.  $e=0$ corresponds to the time period when the policy was implemented.  Negative values of $e$ correspond to pre-treatment estimates of the effect of the policy (and can be thought of as pre-tests), and positive values of $e$ correspond to estimates of the effect of the policy at different lengths of exposure to the treatment.  90\% confidence intervals are provided by the dashed lines. }
    \end{justify}
\end{figure}

Next, we turn to estimates using the approach based on unconfoundedness. As discussed above, these estimates require estimating a model for treatment participation and an outcome regression model. For both of these models, we include a cubic polynomial in the current number of cases per 1000 people in a state (we define the number of current cases as the change in cumulative cases over the previous seven days); we do not include the number of susceptible individuals as this is very close to the full population in all states during the period early in the pandemic that we consider; we additionally include a dummy variable for region of the country so that states are compared to other states in the same region as well as the logarithm of the state's population; finally, we include the 7-day lag of the cumulative cases, the 7-day lag of the number of tests run per 1000 people in the state and the change in tests from the pre-treatment period to the current period in order to control for the possibility that some states were detecting Covid-19 cases better than others. When we estimate the propensity score, we find evidence that the overlap condition is violated indicating that there are a substantial number of states that do not have reasonable comparisons among never-treated and late-treated states. From this step, we drop 14 states from our analysis.\footnote{The states that we drop are Alabama, Arizona, California, Florida, Georgia, Mississippi, Missouri, New York, North Carolina, Pennsylvania, South Carolina, Texas, Virginia, and Washington.}  The omitted states include states such as New York  and California which had large numbers of early Covid-19 cases and were early adopters of SIPOs. It is, therefore, hard to find reasonable comparisons for these states. Another large number of states from the South are dropped due to the timing of their policies being very similar which makes it challenging to find reasonable comparison states when region is included as a conditioning variable.  

These results are provided in \Cref{fig:state-covid-results-unc}.  Using the unconfoundedness approach, we estimate relatively small and statistically insignificant effects of SIPOs.

\subsection{County-Level Results}

There are several complications with any state-level analysis that are worth noting.  First, there are several other Covid-19 policies that were being implemented in a number of states around the same time.  Several recent papers have pointed out challenges with ``controlling for'' other policies in linear models like the one in \Cref{eqn:twfe-es} (e.g., \citet{chaisemartin-dhaultfoeuille-2021b,goldsmith-hull-kolesar-2021}) especially in a framework (like the current one) with treatment effect heterogeneity.  Moreover, using state-level data, the number of observations is very small, and, relatedly, it is challenging to find treated and untreated states that are similar enough to each other to reasonably interpret differences in outcomes as being due to the policy.\footnote{To give a simple specific example, population density is likely to be an important determinant of Covid-19 transmission rates.  One of the largest states that did not implement a SIPO was Arkansas, and a natural strategy is to try to compare Covid-19 cases in Arkansas to, say, Missouri which implemented a SIPO on April 6.  At the state-level, though, the population density of Arkansas and Missouri is much different.  Missouri's population is double Arkansas's population, and it has two major cities, Kansas City and St.\ Louis, while Arkansas does not have a major city.  This suggests that, at the state-level, Arkansas may not be very useful for delivering counterfactual outcomes for Missouri had it not implemented the policy.}  On the other hand, at the county-level, there are often a large number of counties that are located in neighboring states  and have similar characteristics, including both pre-treatment pandemic-related characteristics as well as population, median income, or demographic characteristics.\footnote{There are some potential drawbacks to using county-level data.  One issue is that, by restricting comparisons to counties with similar characteristics, it potentially changes the target parameter.  There are other issues related to inference such as spatial correlations and clustering standard errors at the county-level rather than the state-level as well.  We discuss these issues in more detail in the Supplementary Appendix.}  

Our strategy for the remainder of this section is to use county-level data and compare counties in states that implemented a SIPO to counties in states that did not implement a SIPO while (i) having a similar mix of other policies (both in terms of which types of policies were implemented and their timing) and (ii) making tight comparisons between counties with similar characteristics.  We focus on Arkansas and Iowa which were two states that did not implement a SIPO; we compare them, in turn, to their surrounding states that implemented SIPOs.\footnote{\citet{haynes-kulkarnia-li-siddique-2022} similarly make  comparisons between  counties in states that implemented a SIPO to counties in states that did not.  Compared to our approach, they use TWFE event study regressions and find mixed results with estimates sometimes indicating reductions in Covid-19 cases, sometimes indicating no effect, and sometimes indicating increases in Covid-19 cases.}  Among their surrounding states, Nebraska, Oklahoma, and South Dakota also did not implement a SIPO.  This provides an opportunity to estimate placebo policy effects; that is, to use different estimation strategies in a case where we know that the ``policy effects'' should be equal to 0, which, therefore, provides a way to assess the performance of different estimation strategies.

\subsubsection*{Data}

As for the state-level data, we obtain the number of county-level Covid-19 cases from the CDC COVID Data Tracker and 2019 county-level population from the Census Bureau.  Some of our results below use the number of Covid-19 tests in a particular county which also comes from the CDC COVID Data Tracker.  

We initially include all states that border either Arkansas or Iowa with two exceptions.  We exclude Texas which shares a small border with Arkansas, and we 
include Kansas which ``almost'' borders both Arkansas and Iowa.  Thus, we start with county-level data for thirteen states; these are listed in \Cref{tab:timing}.

\begin{table}[t!]
\centering
\caption{Timing of Covid-Related Policies across States}
\label{tab:timing}
\small
\begin{tabular}{rcccccc}
\toprule
 State & & Emergency Dec. & Schools Closed & Shelter-in-Place & Business Closure & Gathering Rest. \\ \midrule
\multirow{2}{*}{AR} & Start & March 11 & March 16 & - & - & March 27 \\
 & End & - & - & - &  & - \\
\multirow{2}{*}{IL} & Start & March 9 & March 17 & March 21 & March 21 & March 13 \\
& End & - & - & May 30 & - & May 30 \\
\multirow{2}{*}{KS} & Start & March 12 & March 17 & March 30 & - & March 17 \\
& End & - & - & May 4 & - & - \\ 
\multirow{2}{*}{LA} & Start & March 11 & March 13 & March 23 & - & March 13 \\
& End & - & - & May 15 & - & May 15 \\
\multirow{2}{*}{IA} & Start & March 9 & April 3 & - & - & March 17 \\
& End & - & - & - & - & - \\
\multirow{2}{*}{MN} & Start & March 13 & March 18 & March 28 & - & March 28 \\
& End & - & - & May 18 & - & - \\
\multirow{2}{*}{MO} & Start & March 13 & March 23 & April 6 & April 6 & March 23 \\
& End & - & - & May 4 & - & - \\
\multirow{2}{*}{MS} & Start & March 14 & March 19 & April 3 & April 3 & March 24 \\
& End & - & - & April 27 & - & - \\
\multirow{2}{*}{NE} & Start & March 13 & April 1 & - & - & April 3 \\ 
& End & - & May 31 & - & - & - \\
\multirow{2}{*}{OK} & Start & March 15 & March 17 & - & April 1 & March 24 \\
& End & May 11 & - & - & - & May 20 \\
\multirow{2}{*}{SD} & Start & March 13 & March 16 & - & - & April 6 \\
& End & - & April 28 & - & - & May 31 \\
\multirow{2}{*}{TN} & Start & March 12 & March 16 & April 2 & April 1 & March 23 \\
 & End & - & - & May 1 & May 30 & May 30 \\
\multirow{2}{*}{WI} & Start & March 12 & March 18 & March 25 & March 25 & March 17 \\
& End & May 11 & - & May 26 & May 26 & - \\
\bottomrule
\end{tabular}
\begin{justify}
    {\footnotesize \textit{Notes:} The table provides the timing of the main state-level policies that were implemented during the first few months of the pandemic: emergency declarations, school closures, SIPOs, business closures, and gathering restrictions.   The data comes from \citet{fullman-et-al-2021} and uses their classification system of policies.  Policies that were not implemented in a particular state are denoted with a ``-'' in the table.  The end date for policies is provided for polices that were ended before the end of May; policies that did not end by May 31 are listed as ``-'' in the table.  Some policies are somewhat ambiguously defined; this is particularly true for business closures and gathering restrictions for which there is a substantial amount of variation across states in terms of which types of businesses and gatherings were restricted.  For some policies, \citet{fullman-et-al-2021} records multiple instances of the same policy.  In these cases, the table provides the first state-wide, mandated version of the policy, and the end date is the period when there was no longer a state-wide, mandated policy (in some cases, particularly for business closures and gathering restrictions, this can involve overlapping policies with variation in the intensity of the restrictions).  In the Supplementary Appendix, we provide some additional details regarding some particular policies.}
\end{justify}
\end{table}

A major concern for our application is the timing of other Covid-19 related policies across states.  \Cref{tab:timing} provides the date when each state declared an emergency, closed schools, implemented a SIPO, closed non-essential businesses, or imposed gathering restrictions.  Overall, Arkansas tended to implement other policies with very similar timing as all of its surrounding states especially with respect to the timing of the emergency declaration and school closures.  There is more variation with respect to non-essential business closures and gathering restrictions though we note that these policies are less clearly defined than the other main Covid-related policies.  Our interpretation is that it is reasonable to consider the mix and timing of other policies as being quite similar to all of its neighboring states.  The timing of Iowa's policies are somewhat more different from its surrounding states.  The primary difference is that Iowa closed its schools somewhat later than surrounding states with the exception of Nebraska.  This arguably suggests that the results below that use Iowa as the comparison state are somewhat less credible (with a viable alternative explanation that differences in school closure policy could be at least partially driving the results).

A second major concern is being able to find counties with similar pre-SIPO pandemic related characteristics.  For each pair of states, we use the same model for the outcome regression and propensity score as for the state-level data that includes a cubic polynomial in the current number of cases, the 7-day lag of cumulative cases, the 7-day lag and change in the number of Covid-19 tests in the county, and the logarithm of county population.  In order to enforce the overlap condition between treated and untreated states, we drop treated counties with an estimated propensity score greater than 0.95.\footnote{In practice, for our main results, we do not drop many counties from this procedure.  When we compare Missouri to Arkansas, we do not drop any counties using this criteria.  When we compare Minnesota to Iowa, we drop Ramsey County which is the county where St.\ Paul is located.  As discussed above, dropping counties this way does change the target parameter from being the $ATT$ to being an $ATT$-type parameter that is local to the region of common support.  And, for example, our approach tends to drop urban counties relative to suburban or rural counties; if SIPOs tended to reduce Covid-19 cases more in urban counties than in other counties, then our estimation procedure would tend to result in smaller in magnitude estimates of policy effects.}  With the remaining set of counties, including both treated and untreated counties, we compute the propensity score weights from \Cref{thm:unc} using the same set of covariates as mentioned above.  Given these weights, we compute the effective sample size for untreated states, which is defined as $\sum_{l \in \mathcal{U}} w_l^2 / \Big(\sum_{l \in \mathcal{U}} w_l^2\Big)$ where $\mathcal{U}$ denotes the set of untreated county (see, for example, \citet{kish-1965,chattopadhyay-zubizarreta-2022,shook-hudgens-2022}).  Then, we exclude combinations of states where the effective sample size is less than or equal to 25 counties for either the treated or untreated state.  Following this procedure, we end up with four combinations of states that satisfy this criteria.  Two of these, (i) Oklahoma relative to Arkansas and (ii) South Dakota relative to Iowa, arise for states that did not implement a SIPO.  We estimate placebo policy effects for these states.  The other two combinations of states are (iii) Missouri relative to Arkansas and (iv) Minnesota relative to Iowa.  These states provide a chance to estimate effects of policies that were actually implemented.  

\subsubsection*{Results}

\begin{figure}[t]
    \centering
    \caption{Placebo Policy Effects}
    \label{fig:county-placebo}
    \begin{subfigure}[b]{.42\textwidth}
    \includegraphics[width=\textwidth]{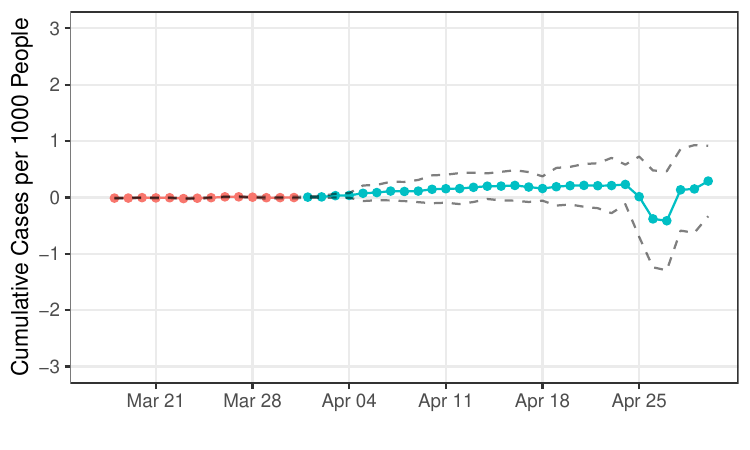}
    \caption{Unconfoundedness: Oklahoma and Arkansas}
    \end{subfigure}
    \begin{subfigure}[b]{.42\textwidth}
    \includegraphics[width=\textwidth]{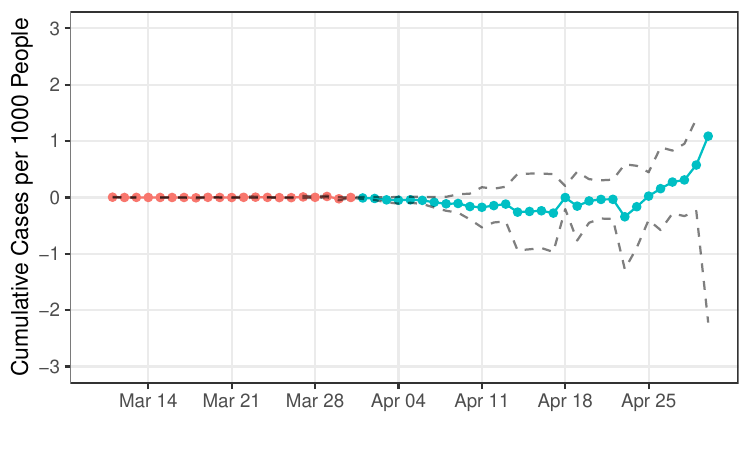}
    \caption{Unconfoundedness: South Dakota and Iowa}
    \end{subfigure}
    \begin{subfigure}[b]{.42\textwidth}
    \includegraphics[width=\textwidth]{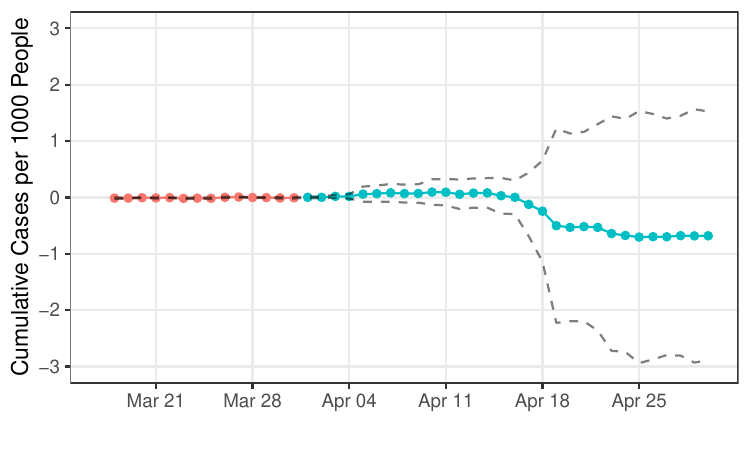}
    \caption{DID: Oklahoma and Arkansas}
    \end{subfigure}
    \begin{subfigure}[b]{.42\textwidth}
    \includegraphics[width=\textwidth]{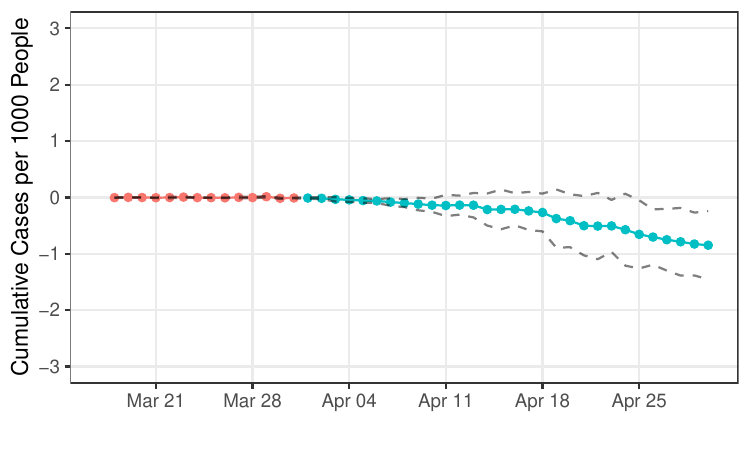}
    \caption{DID: South Dakota and Iowa}
    \end{subfigure}
    
    \begin{justify}
    { \footnotesize \textit{Notes:} The figure provides estimates of placebo SIPOs (i.e., policies that were not actually implemented)  for Oklahoma and South Dakota using the unconfoundedness and DID approaches discussed in the main text.  The placebo policy date is set to be April 1 for both Oklahoma and South Dakota.  Estimates before the placebo policy date are in red while estimates in periods after the placebo policy date are provided in blue.  The dashed lines provide 90\% pointwise confidence intervals.}
    \end{justify}
\end{figure}

We start by providing placebo policy effect estimates for Oklahoma and South Dakota --- neither of which implemented a SIPO.  The reason for starting here is that this serves as a good way to evaluate estimation strategies; policy effects should, in principle, be equal to 0 in all time periods.  In order to facilitate comparisons to other results, for states that did not implement a SIPO, we use a ``placebo policy'' date of April 1.  These results are provided in \Cref{fig:county-placebo}.  Panels (a) and (b) provide estimates using the unconfoundedness approach.  These estimates are generally small and not statistically significantly different from zero.  Estimates using a difference-in-differences approach are provided in panels (c) and (d).  For Oklahoma, none of the post-policy estimates are statistically different from zero though the standard errors are notably larger using DID than for the unconfoundedness approach.  More notably, using DID, we spuriously estimate that the placebo policy in South Dakota reduced the number of Covid-19 cases in South Dakota.  These results suggest that, at least for these two policies, the unconfoundedness approach performs better than DID. It is also worth mentioning that, due to the relatively similar pre-pandemic characteristics of Oklahoma to Arkansas and South Dakota to Iowa, this is still a relatively favorable setting for DID estimation strategies.

\begin{figure}[t]
    \centering
    \caption{County-Level Estimates under Unconfoundedness}
    \label{fig:county-main-results}
    \begin{subfigure}[b]{.42\textwidth}
    \includegraphics[width=\textwidth]{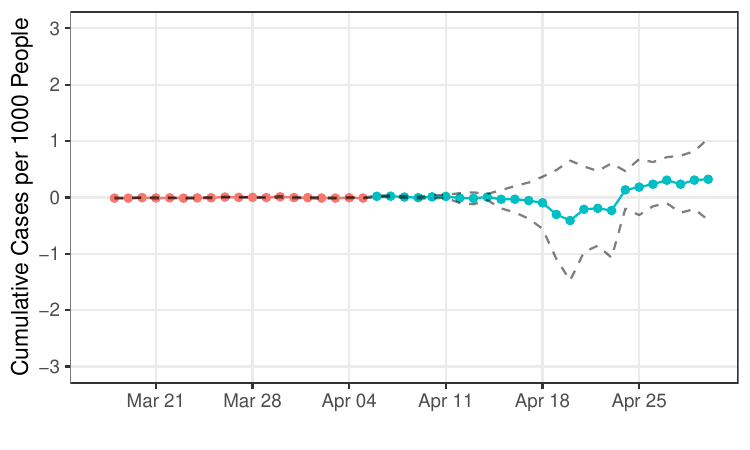}
    \caption{Missouri and Arkansas}
    \end{subfigure}
    \begin{subfigure}[b]{.42\textwidth}
    \includegraphics[width=\textwidth]{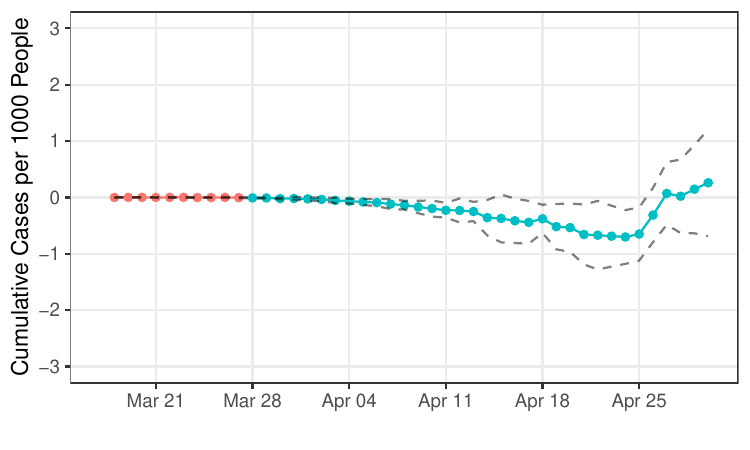}
    \caption{Minnesota and Iowa}
    \end{subfigure}
    \begin{justify}
    { \footnotesize \textit{Notes:} The figure provides estimates of SIPOs effects on Covid-19 cases for Missouri and Minnesota using the unconfoundedness approach discussed in the main text.  Pre-treatment estimates are reported in red while post-treatment estimates are reported in blue.  The pre-treatment estimates use the immediately preceding day as the base period while all the post-treatment periods use the period immediately before the treatment as the base period.  The dashed lines provide 90\% pointwise confidence intervals.}
    \end{justify}
\end{figure}

Next, we move to estimating effects of policies that were actually implemented in Missouri (using counties from Arkansas as the comparison) and in Minnesota (using counties from Iowa as the comparison).  The results under unconfoundedness are provided in \Cref{fig:county-main-results}.     For Missouri, we do not find evidence of policy effects on Covid-19 cases.  On the other hand, for Minnesota, we estimate that its SIPO decreased Covid-19 cases at least in some periods.  For example, on April 25, we estimate that Minnesota's SIPO reduced Covid-19 cases by about 0.46 cases per 1000 people in the state.  In results provided in the 
Supplementary Appendix, using DID, we estimate a large reduction in Covid-19 cases in Missouri due to the policy though the standard errors are much larger than for the results based on unconfoundedness and not statistically different from 0.  For Minnesota, the DID estimates are quite similar to the unconfoundedness results presented here.  

In the Supplementary Appendix, we provide additional estimates for all combinations of states discussed in this section.  Of the remaining 8 combinations of states that (i) implemented a policy and (ii) exhibit similar pre-treatment paths of Covid-19 cases, using the same DID estimation strategy used in this section, two of the estimates are positive and statistically significant, two of the estimates are negative and statistically significant, and four of the estimates are not statistically different from zero.  Thus, like the case with state-level data above, using DID can lead to hard-to-explain positive estimates of SIPOs on Covid-19 cases and, more generally, estimates that seem inconsistent with each other. Arguably, these differences could be explained by heterogeneous effects of policies implemented in different states, though, in our view, the large magnitude of the differences in estimated policy effects cuts against heterogeneous policy effects being a full explanation of the differences.  Instead, a better explanation seems to be that these differences are driven by pre-treatment differences in the state of the pandemic between counties in states that implemented SIPOs relative to counties in neighboring states that did not.

\begin{figure}[t!]
    \centering
    \caption{Estimates of SIPO Orders on Travel}
    \label{fig:travel-results}
    \begin{subfigure}[b]{.49\textwidth}
    \includegraphics[width=\textwidth]{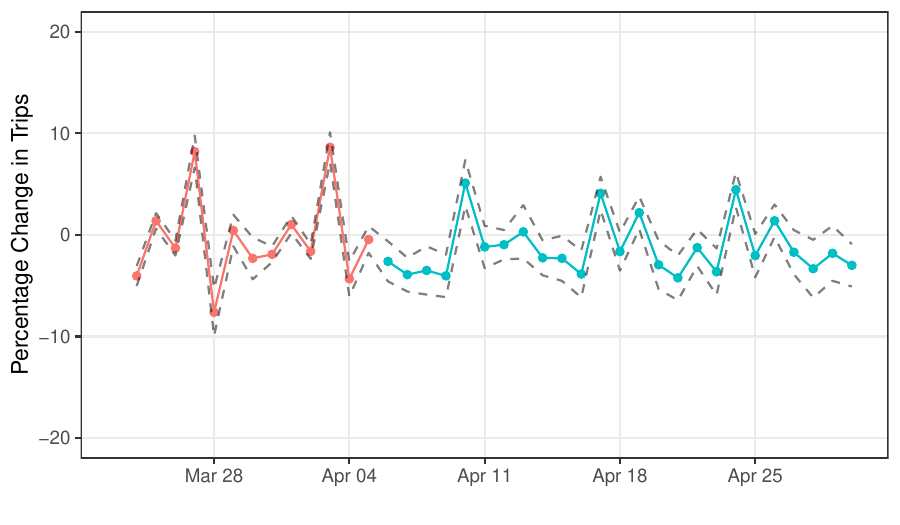}
    \caption{Standard DID}
    \end{subfigure}
    \begin{subfigure}[b]{.49\textwidth}
    \includegraphics[width=\textwidth]{Missouri_Arkansas_oo_did.pdf}
    \caption{Regression DID}
    \end{subfigure}
    \begin{subfigure}[b]{.49\textwidth}
    \includegraphics[width=\textwidth]{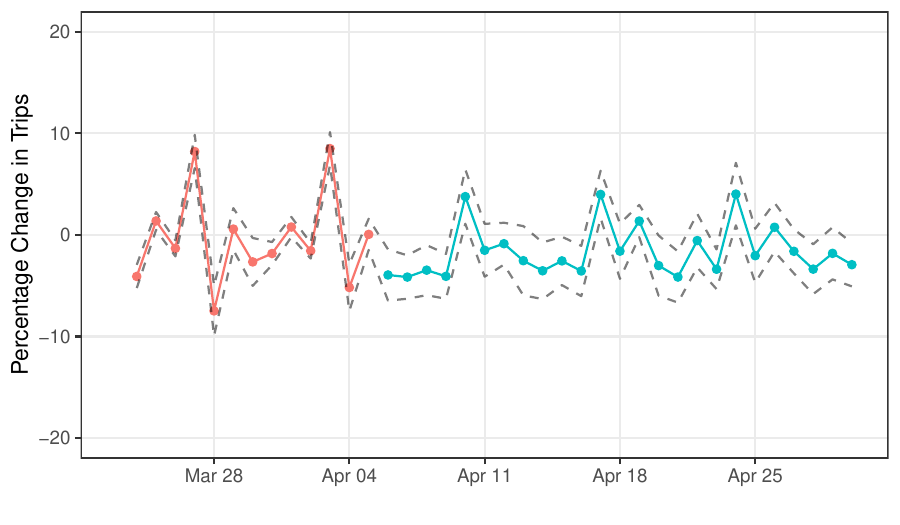}x
    \caption{Adjusted Regression DID}
    \end{subfigure}
    
    \begin{justify}
    { \footnotesize \textit{Notes:}  The figure contains event study type estimates of the effect of SIPOs on the percentage change in retail and recreation travel.  $e=0$ corresponds to the time period when the policy was implemented.  Negative values of $e$ correspond to pre-treatment estimates of the effect of the policy and can be thought of as pre-tests, and positive values of $e$ correspond to estimates of the effect of the policy at different lengths of exposure to the treatment.  Panel (a) provides estimates using standard DID (without accounting for cases), Panel (b) provides regression DID estimates (accounting for cases but not that the policy may have a direct effect on cases), and Panel (c) provides adjusted regression DID estimates (accounting for cases and allowing for the policy to have had an effect on cases as is proposed in the text).  The dashed line provides 90\% pointwise confidence intervals.}
    \end{justify}
\end{figure}

Finally, we consider the effect of SIPOs on travel.  We focus on the percentage change in retail and recreation travel from a pre-Covid baseline.  In the main text, we focus on the effect of the policy on travel in Missouri using Arkansas as the comparison state.%
These results are available in \Cref{fig:travel-results}.  The results in Panel (a) come from a standard DID approach that implicitly imposes that Covid-19 cases do not directly affect travel; the results in Panel (b) come from the regression DID approach that includes current cases as a covariate but not accounting for the possibility that SIPOs could have affected the number of cases directly; and the results in Panel (c) use the adjusted regression DID approach proposed in the current paper that allows for the policy to have had an effect on Covid-19 cases.

In this case, the estimates are more broadly similar than they were for cumulative Covid-19 cases.  Standard DID estimates indicate a relatively small but persistent negative effect of SIPOs on retail and recreation travel.  Using standard DID, the overall estimate of the effect of SIPOs across post-treatment time periods is -1.30 (p-value: 0.16); i.e., across the first several weeks of the SIPO, the policy reduced travel by about 1.3 percent relative to what travel would have been if the policy had not been implemented.  The point estimates from regression DID (these estimates just include observed current cases as a covariate) are somewhat larger in magnitude; in this case, the estimated overall effect of SIPOs on travel is -1.63 (p-value: 0.08).  Finally, the adjusted regression DID overall estimated effect of SIPOs  is -1.25 (p-value: 0.19) when we allow for the policy to have had an effect on cases.  In the Supplementary Appendix, we provide analogous results using the state-level data from earlier in this section.  Those results are broadly similar to the ones presented here though we estimate somewhat larger-in-magnitude effects (around a 3 percent reduction in travel due to SIPOs) that do not vary much by estimation strategy and are somewhat more precisely estimated than the results presented here.  

\subsection{Discussion of Results}

Our results, especially those about the effect of SIPOs on the number of Covid-19 cases, are substantially different from existing estimates, and it is worth making a few additional comments.  First, our results on the effect of SIPOs on travel are more similar to existing estimates (e.g., \citet{goolsbee-syverson-2021}).  Since the primary channel through which SIPOs would likely reduce Covid-19 cases is through reducing travel/contact with other individuals, it seems reasonable to simultaneously estimate relatively small effects of SIPOs on travel coinciding with small effects of SIPOs on the number of Covid-19 cases; but harder to rationalize SIPOs strongly decreasing Covid-19 cases while having only a small effect on travel.

That being said, we hesitate to interpret our results as providing strong evidence that SIPOs did not reduce Covid-19 cases.  For one thing, the interpretation of our treatment effect parameters is somewhat subtle.  Untreated potential outcomes here do not correspond to particular locations not reacting at all to Covid-19 but rather to outcomes that would have occurred if a state had not implemented the policy (but other things about the state remained the same).  This can be seen to be clearly relevant from the summary statistics in \Cref{tab:ss} where all states (not just those who implemented the policy) were experiencing massive decreases in travel over the period that we consider.   Importantly, this indicates that our results are not at all saying that staying at home did not have an effect on Covid-19 cases.  Second, the statistical power of all of the estimation procedures considered above is relatively low (this is true both for the unconfoundedness approach and the DID approaches) and leads to generally wide confidence intervals on estimated policy effects.  And even though a number of our estimates are not statistically significant, many of them  are large enough to be compatible with a wide variety of possible effects of SIPOs on Covid-19 cases.\footnote{To give an example, in our results for Missouri, which is a state for which we do not estimate a statistically significant policy effects on Covid-19 cases in post-treatment periods (see \Cref{fig:county-main-results}), the lower end of a 90\% confidence interval for our estimated effect of the policy on Covid-19 cases 21 days after the policy was implemented is that it reduced Covid-19 cases by about 0.13 per 1000 people.  If we divide this by the estimated number of Covid-19 cases that treated locations would have experienced in the same period if they had not implemented the policy (i.e., $\E[C_t(0)|D=1]$ in \Cref{eqn:attc} which is identified and can be recovered in our setup), we would estimate that SIPOs decreased Covid-19 cases by 30\%.  In other words, our estimates do not necessarily rule out the possibility that SIPOs may have had quite large effects on Covid-19 cases.}  Instead, we interpret the results from the application as indicating that these sorts of policies are likely to be very challenging to precisely evaluate due to both data limitations as well as trying to deal with a highly nonlinear outcome and a policy that was adopted at different times by states whose exposure to the pandemic also varied widely and was correlated with the timing of the policy being adopted.

\FloatBarrier

\section{Conclusion}

In this paper, we have considered several different policy evaluation strategies and how compatible they are with a leading epidemic model.   For identifying the direct effects of policies on the number of Covid-19 cases, our results suggest that strategies based on unconfoundedness type assumptions are likely to perform better than difference-in-differences type strategies due to the highly nonlinear nature of the spread of Covid-19.  

Our second main set of results were about evaluating the effects of policies on other economic outcomes when (i) the policy can affect the number of Covid-19 cases and (ii) the number of Covid-19 cases can have a direct effect on the economic outcome of interest.  For this case, we also showed that two of the most common ways to evaluate these policies (difference-in-differences directly or including the number of cases as a covariate in a DID setup) do not generally deliver an average effect of the policy.  We proposed an alternative estimator that is valid in this case.

We applied our approach to study the effects of shelter-in-place orders on Covid-19 cases and travel early in the pandemic.  We showed that our theoretical arguments were indeed relevant in this context and led to notably different estimates (particularly for the number of Covid-19 cases) relative to the most common approaches used in applications.

There remain a number of interesting possible extensions to this work, and we conclude by mentioning two of them.  First, evaluating the effects of various policies (especially early in the pandemic) is complicated by limited and nonrandom Covid-19 testing during the early part of the pandemic (see, for example, \citet{callaway-li-2021,manski-molinari-2020}), and it would be interesting to extend our results along these dimensions.  Second, another common policy evaluation approach in the context of Covid-19 related policies is the synthetic control method (examples include \citet{cho-2020,dave-friedson-matsuzawa-mcnichols-sabia-2020,friedson-mchnichols-sabia-dave-2021,mitze-kosfeld-rode-walde-2020}).  It appears that, most often, a researcher's decision between difference-in-differences or synthetic controls is driven by whether the number of treated locations is large or small.  However, it is less clear under what conditions synthetic control approaches are compatible with the sorts of epidemiological models that we considered in the current paper.

\bigskip

\bigskip

\singlespacing

\printbibliography

\appendix

\numberwithin{equation}{section}

\FloatBarrier

\section{More Details on Stochastic SIRD Models} \label{app:sim-details}

\subsection{Stochastic SIRD Model for Untreated Potential Outcomes} \label{sec:sird-for-untreated}

In this section, we write down a stochastic SIRD model along the lines of Equations \ref{eqn:transition-infections}-\ref{eqn:transition-total-cases} but for untreated potential outcomes and written in an error form.

\begin{namedassumption}{Stochastic SIRD Model for Untreated Potential Outcomes} \label{ass:sird-model} For all $t=2, \ldots, \mathcal{T}$
\begin{align}
    I_{lt}(0) &= I_{lt-1}(0) + \left( \beta \frac{I_{lt-1}(0)}{N_l} S_{lt-1}(0) + u^{I}_{lt} \right) - \Delta R_{lt}(0) - \Delta \delta_{lt}(0) \label{eqn:sird0I}\\
    R_{lt}(0) &= R_{lt-1}(0) + \lambda I_{lt-1}(0) + u^{R}_{lt} \label{eqn:sird0R}\\
    \delta_{lt}(0) &= \delta_{lt-1}(0) + \gamma I_{lt-1}(0) + u^{\delta}_{lt} \label{eqn:sird0D}\\
    N_l &= S_{lt}(0) + I_{lt}(0) + R_{lt}(0) + \delta_{lt}(0) \label{eqn:sird0N}\\
    C_{lt}(0) &= N_l - S_{lt}(0) \label{eqn:sird0C}
\end{align}
where we omit an equation for $S_{lt}(0)$, the number of susceptible in location $l$ in time period $t$, because it is fully determined by $I_{lt}(0),R_{lt}(0),\delta_{lt}(0)$, and $N_l$.  Next, define $u_{lt} := (u_{lt}^I, u_{lt}^R, u_{lt}^\delta)'$.  We also make the following assumptions about $u_{lt}$
\begin{align}
    \E[u_t|\mathcal{F}_{t-1}(0), \ldots, \mathcal{F}_{1}(0), D=d]=0 \textrm{ for all $t=2,\ldots,\mathcal{T}$} \label{eqn:ut-ass1}
\end{align}
and that
\begin{align}
    u_t \independent (D, \mathcal{F}_{t-2}(0), \ldots, \mathcal{F}_{1}(0)) | \mathcal{F}_{t-1}(0) \textrm{ for all $t=2,\ldots,\mathcal{T}$} \label{eqn:ut-ass2}
\end{align}
\end{namedassumption}
Together with Equations \ref{eqn:sird0I}-\ref{eqn:sird0C}, \Cref{eqn:ut-ass1} completes the stochastic SIRD model for untreated potential outcomes in an analogous way to the model presented in the main text.  Note that, together, these conditions rule out that $u_t$ are serially correlated.  Additionally, \Cref{eqn:ut-ass1,eqn:ut-ass2} are related to the Markov property of stochastic SIRD models discussed in the main text in that the distribution of error terms can depend on the state of the pandemic in the immediately preceding period (this allows, for example, the variance of the error terms to depend on the state of the pandemic in the previous period), but their distribution does not depend on further lags of the state of the pandemic.  These conditions also have the flavor of sequential exogeneity assumptions in the panel data econometrics literature where the distribution of error terms does not depend on lags of the state of the pandemic (besides the first one), but current shocks can and do impact future pandemic states.  Both \Cref{eqn:ut-ass1,eqn:ut-ass2} also involve error terms being independent of treatment status conditional on pre-treatment status.  This is arguably the key condition here and indicates that, given the state of the pandemic in period $t-1$, the distribution of shocks for treated locations is the same as for untreated locations.  Relative to typical implementations of stochastic SIRD models, these are  mild conditions.  For example, it is typical to impose that, $\left( \beta \frac{I_{lt-1}(0)}{N_l} S_{lt-1}(0) + u^I_{lt} \right)$, which is the number of new infections, follows a Poisson distribution with parameter $\beta \frac{I_{lt-1}(0)}{N_l} S_{lt-1}(0)$, and that the number of new recoveries ($\lambda I_{lt-1}(0) + u^R_{lt}$), new deaths ($\gamma I_{lt-1}(0) + u^\delta_{lt}$), and continued infections follows a multinomial distribution with parameters $(\lambda I_{lt-1(0)}, \gamma I_{lt-1}(0), (1-\lambda-\gamma) I_{lt-1}(0))$.  This set up is compatible with the \Cref{ass:sird-model} discussed above.

\subsection{Simulation Details}

\Cref{tab:sim-params} provides the values of the parameters that we use in the simulations.  The most important things to notice are that (i) the policy has no effect on Covid-19 infections rates (i.e., $\beta=\beta_{pol}$) and (ii) the distribution of the timing of the first case is different across locations that participate in the treatment and those that do not.  This is roughly analogous to the idea that locations that were exposed to Covid-19 earlier tended to implement policies earlier.

\begin{table}[ht]
\centering
\caption{Simulation Parameters}
\begin{tabular}{lcr} \label{tab:sim-params}
parameter & notation & value \\
\hline
infection rate                  & $\beta$       & 0.08                            \\
post policy $\beta$             & $\beta_{pol}$ & 0.08                           \\
recovery rate                   & $\lambda$      & 0.04                            \\
death rate                      & $\gamma$      & 0.003                           \\
time periods                    & $T$           & 400                             \\
population size                 & $N$           & 1000                             \\
treatment probability           & $p$           & 0.5 \\
initial case period (treated)   &               & $\textrm{Poisson}(\lambda=40)$  \\
initial case period (untreated) &               & $\textrm{Poisson}(\lambda=80)$  \\
initial \# cases                &               & 10                              \\
\# locations                    &               & 250                            \\
policy start period & & 150
\end{tabular}
\end{table}

\section{Proofs} \label{sec:proofs}

\subsection*{Proof of \Cref{thm:did-bias}}
\begin{proof}
    For the first part, notice that
    \begin{align}
        \E[\Delta^{(t^*-1, t)} C_t(0) | D=d] &= \E[C_t(0) - C_{t^*-1}(0) | D=d] \nonumber \\
        &= \sum_{s=t^*}^{t} \E[\Delta C_s(0) | D=d] \nonumber \\
        &= \sum_{s=t^*}^{t} \E\Big[ \E[\Delta C_s(0) | \mathcal{F}_{t^*-1}, D=d] | D=d\Big] \label{eqn:did-path}
    \end{align}
    where the first equality holds by the definition of $\Delta^{(t^*-1,t)}C_t(0)$, the second equality holds by adding and subtracting $\E[C_s(0)|D=d]$ for all $s = t^*, \ldots, (t-1)$, and the last equality holds by the law of iterated expectations.

  For the second part, 
  \begin{align*}
      ATT^{C}_t &= \E[C_t(1) - C_t(0) | D=1] \\
      &= \E[C_t(1) - C_{t^*-1}(0) | D=1] - \E[ \Delta^{(t^*-1,t)} C_t(0) | D=1] \\
      &= \E[C_t(1) - C_{t^*-1}(0) | D=1] -  \E[ \Delta^{(t^*-1,t)} C_t(0) | D=0] \\ 
      & \hspace{25pt} + \Big(\E[ \Delta^{(t^*-1,t)} C_t(0) | D=0] - \E[ \Delta^{(t^*-1,t)} C_t(0) | D=1] \Big) \\
      &= \E[\Delta^{(t^*-1,t)} C_t | D=1] - \E[\Delta^{(t^*-1,t)} C_t | D=0] \\
      & \hspace{25pt} + \left(\sum_{s=t^*}^{t} \E\Big[ \E[\Delta C_s(0) | \mathcal{F}_{t^*-1}, D=0] | D=0\Big] - \sum_{s=t^*}^{t} \E\Big[ \E[\Delta C_s(0) | \mathcal{F}_{t^*-1}, D=1] | D=1\Big]\right)
  \end{align*}
  where the first equality comes from the definition of $ATT^C_t$, the second equality holds by adding and subtracting $\E[C_{t^*-1}(0)|D=1]$ (which is the average number of Covid-19 cases across treated locations in the pre-treatment period), the third equality holds by adding and subtracting $\E[\Delta^{(t^*-1,t)}C_t(0)|D=0]$ (which is the average path of Covid-19 cases from period $t^*-1$ to $t$ among untreated locations), and the fourth equality holds by rewriting potential outcomes in terms of their observed counterparts and by \Cref{eqn:did-path}.  This implies the second result in the theorem.
\end{proof}

\subsection*{Proof of \Cref{prop:unc}}

To conserve on notation, define the function $r(\mathcal{F}_{t}(0), u_{t})$ to be the vector-valued function where the first argument takes in the state of the pandemic in some period $t$, $\mathcal{F}_{t}(0)$, and vector of error terms $u_{t}$ and returns $\mathcal{F}_{t+1}(0)$ according to the \Cref{ass:sird-model}.  Then, for some post-treatment time period $t \geq t^*$, we have that
\begin{align}
    \mathcal{F}_{lt} &= r(\mathcal{F}_{lt-1}, u_{lt}) \label{eqn:r}
\end{align}

Before proving \Cref{prop:unc}, we provide two helpful intermediate results. 

\begin{lemma} \label{lem:1}
    In the \Cref{ass:sird-model} and under \Cref{ass:overlap}, for any $t \geq t^*$, for a function $h$, $d \in \{0,1\}$, and for some $k>0$ such that $t-k\geq 1$,
    \begin{align*}
        \E[h(\mathcal{F}_t(0))|\mathcal{F}_{t-1}(0), \ldots, \mathcal{F}_{t-k}(0), D=d] = \E[h(\mathcal{F}_t(0))|\mathcal{F}_{t-1}(0), D=d]
    \end{align*}
\end{lemma}

\begin{proof}
    Notice that
    \begin{align*}
        \E[h(\mathcal{F}_t(0))|\mathcal{F}_{t-1}(0), \ldots, \mathcal{F}_{t-k}(0), D=d] &= \E[h(r(\mathcal{F}_{t-1}(0),u_t))|\mathcal{F}_{t-1}(0), \ldots, \mathcal{F}_{t-k}, D=d] \\
        &= \E[h(r(\mathcal{F}_{t-1}(0),u_t))|\mathcal{F}_{t-1}(0), D=d] \\
        &= \E[h(\mathcal{F}_{t}(0))|\mathcal{F}_{t-1}(0), D=d]
    \end{align*}
    where the first and last equalities holds by the definition of $r$, and the second equality holds by \Cref{eqn:ut-ass2} in the \Cref{ass:sird-model}.
\end{proof}

\begin{lemma} \label{lem:2}
    In the \Cref{ass:sird-model} and under \Cref{ass:overlap}, for any $t \geq t^*$ and for a function $h$,
    \begin{align*}
        \E[h(\mathcal{F}_t(0))|\mathcal{F}_{t-1}(0), D=1] = \E[h(\mathcal{F}_t(0))|\mathcal{F}_{t-1}(0), D=0]
    \end{align*}
\end{lemma}

\begin{proof}
    Notice that
    \begin{align*}
        \E[h(\mathcal{F}_t(0))|\mathcal{F}_{t-1}(0), D=1] &= \E[h(r(\mathcal{F}_{t-1}(0),u_t))|\mathcal{F}_{t-1}(0), D=1] \\
        &= \E[h(r(\mathcal{F}_{t-1}(0),u_t))|\mathcal{F}_{t-1}(0), D=0] \\
        &=\E[h(\mathcal{F}_t(0))|\mathcal{F}_{t-1}(0), D=0]
    \end{align*}
    where the first and last equalities hold by the definition of $r$ and the second equality holds by \Cref{eqn:ut-ass2} in the \Cref{ass:sird-model}.
\end{proof}

\begin{proof}[Proof of \Cref{prop:unc}]

For the proof, we provide the more general result that $\E[\mathcal{F}_t(0)|\mathcal{F}_{t^*-1}(0), D=1] = \E[\mathcal{F}_t(0)|\mathcal{F}_{t^*-1}(0), D=0]$; i.e., we prove unconfoundedness holds for all the variables in the epidemic model which includes cumulative cases, but also includes the number of susceptible, the number of currently infected, and the number of recovered and deaths.

For $k \geq 1$, we recursively define
\begin{align*}
    g_d^{(k)}(\mathcal{F}_{s-k}(0)) := \begin{cases} \E[g_d^{(k-1)}(\mathcal{F}_{s-k+1}(0))|\mathcal{F}_{s-k}(0), D=d] & \textrm{ if } k > 1 \\
    \E[\mathcal{F}_{s}(0) | \mathcal{F}_{s-1}(0), D=d] & \textrm{ if } k = 1\end{cases}
\end{align*}
Using this notation, notice that we can write
\begin{align}
    \E[\mathcal{F}_t(0) | \mathcal{F}_{t-1}(0), D=1] = g^{(1)}_1(\mathcal{F}_{t-1}(0)) \label{eqn:proof1}
\end{align}
and that 
\begin{align}
    \E[\mathcal{F}_t(0) | \mathcal{F}_{t-2}(0), D=1] &= \E\Big[ \E[\mathcal{F}_{t}(0) | \mathcal{F}_{t-1}(0), \mathcal{F}_{t-2}(0), D=1] \Big| \mathcal{F}_{t-2}(0), D=1\Big] \nonumber \\
    &= \E\Big[ \E[\mathcal{F}_{t}(0) | \mathcal{F}_{t-1}(0), D=1] \Big| \mathcal{F}_{t-2}(0), D=1\Big] \nonumber \\
    &= \E\Big[ g^{(1)}_1(\mathcal{F}_{t-1}(0)) \Big| \mathcal{F}_{t-2}(0), D=1\Big] \label{eqn:proof2} \\
    &= g^{(2)}_1(\mathcal{F}_{t-2}(0)) \label{eqn:proof3}
\end{align}
where the first equality holds by the law of iterated expectations, the second equality by \Cref{lem:1}, the third equality by the definition of $g_1^{(1)}$, and the last equality by the definition of $g^{(2)}_1$.  Similarly, notice that
\begin{align*}
    \E[\mathcal{F}_t(0) | \mathcal{F}_{t-3}(0), D=1] &= \E\Big[ \E[\mathcal{F}_{t}(0) | \mathcal{F}_{t-1}(0), \mathcal{F}_{t-2}(0), \mathcal{F}_{t-3}(0), D=1] \Big| \mathcal{F}_{t-3}(0), D=1\Big] \\
    &= \E\Big[ \E[\mathcal{F}_{t}(0) | \mathcal{F}_{t-1}(0), D=1] \Big| \mathcal{F}_{t-3}(0), D=1\Big] \\
    &= \E\Big[ g^{(1)}_1(\mathcal{F}_{t-1}(0)) \Big| \mathcal{F}_{t-3}(0), D=1\Big] \\
    &= \E\Big[ \E[g^{(1)}_1(\mathcal{F}_{t-1}(0)) | \mathcal{F}_{t-2}(0), \mathcal{F}_{t-3}(0), D=1] \Big| \mathcal{F}_{t-3}(0), D=1\Big] \\
    &= \E\Big[ \E[g^{(1)}_1(\mathcal{F}_{t-1}(0)) | \mathcal{F}_{t-2}(0), D=1] \Big| \mathcal{F}_{t-3}(0), D=1\Big] \\
    &= \E\Big[ g^{(2)}_1(\mathcal{F}_{t-2}(0)) \Big| \mathcal{F}_{t-3}(0), D=1\Big] \\
    &= g^{(3)}_1(\mathcal{F}_{t-3}(0))
\end{align*}
where the first equality holds by the law of iterated expectations, the second equality holds by \Cref{lem:1}, the third equality holds by the definition of $g^{(1)}_1$, the fourth equality holds by the law of iterated expectations, the fifth equality holds by \Cref{lem:1}, the sixth equality holds by the definition of $g^{(2)}_1$, and the last equality holds by the definition of $g^{(3)}_1$.  Continuing to proceed in this way, we have that
\begin{align*}
    \E[\mathcal{F}_t(0)|\mathcal{F}_{t^*-1}(0), D=1] &= \E\Big[ g^{(t-t^*)}_1(\mathcal{F}_{t^*}(0)) \Big| \mathcal{F}_{t^*-1}(0), D=1\Big] \\
    &= g^{(t-t^*+1)}_1(\mathcal{F}_{t^*-1}(0))
\end{align*}
The same sorts of arguments can be used to show that $\E[\mathcal{F}_t(0)|\mathcal{F}_{t^*-1}(0), D=0] = g^{(t-t^*+1)}_0(\mathcal{F}_{t^*-1}(0))$.

Next, notice that \Cref{lem:2}, together with \Cref{eqn:proof1} implies that
\begin{align}
    g^{(1)}_1(\mathcal{F}_{t-1}(0)) = \E[\mathcal{F}_t(0)|\mathcal{F}_{t-1}(0), D=1] = \E[\mathcal{F}_t(0)|\mathcal{F}_{t-1}(0), D=0] = g^{(1)}_0(\mathcal{F}_{t-1}(0)) \label{eqn:proof4}
\end{align}
Similarly,
\begin{align*}
    g^{(2)}_1(\mathcal{F}_{t-2}(0)) &= \E\Big[ g^{(1)}_1(\mathcal{F}_{t-1}(0)) \Big| \mathcal{F}_{t-2}(0), D=1\Big] \\
    &= \E\Big[ g^{(1)}_1(\mathcal{F}_{t-1}(0)) \Big| \mathcal{F}_{t-2}(0), D=0\Big] \\
    &= \E\Big[ g^{(1)}_0(\mathcal{F}_{t-1}(0)) \Big| \mathcal{F}_{t-2}(0), D=0\Big] \\
    &= g^{(2)}_0(\mathcal{F}_{t-2}(0))
\end{align*}
where the first and last equalities hold by the definition of $g^{(2)}_d$, the second equality holds by \Cref{lem:2}, and the third equality holds by \Cref{eqn:proof4}.

Proceeding along these lines, it follows that 
\begin{align*} 
g^{(k)}_1(\mathcal{F}_{t-k}(0)) = g^{(k)}_0(\mathcal{F}_{t-k}(0)) \textrm{ for all $k=1, \ldots, (t-t^*+1)$}
\end{align*}
Taking $k=t-t^*+1$ implies that $\E[\mathcal{F}_t(0)|\mathcal{F}_{t^*-1}(0), D=1] = \E[\mathcal{F}_t(0)|\mathcal{F}_{t^*-1}(0), D=0]$ which completes the proof.

\end{proof}

\subsection*{Proof of \Cref{thm:unc}}

\begin{proof}
    Given the result in \Cref{prop:unc}, the proof of this result holds under standard arguments for unconfoundedness.  We provide them here for completeness.  First, notice that
    \begin{align}
        ATT^C_t &= \E[C_t(1) - C_t(0) | D=1] \nonumber \\
        &= \E[C_t | D=1] - \E\left[ \E[C_t | \mathcal{F}_{t^*-1}, D=0] | D=1 \right] \label{eqn:attc-1}
    \end{align}
    where the first equality holds from the definition of $ATT^C_t$, and the second equality holds by \Cref{prop:unc}.  This completes the first part of the proof.  For the second part, notice that continuing from \Cref{eqn:attc-1},
    \begin{align}
        ATT^C_t &= \E\left[ \frac{D}{\E[D]} C_t \right] - \E\left[ \frac{\E[(1-D)C_t | \mathcal{F}_{t^*-1} ] }{1-p(\mathcal{F}_{t^*-1})} \Big| D=1\right] \nonumber \\
        &= \E\left[ \frac{D}{\E[D]} C_t \right] - \E\left[ \frac{p(\mathcal{F}_{t^*-1})} {\E[D](1-p(\mathcal{F}_{t^*-1}))} \E[(1-D)C_t | \mathcal{F}_{t^*-1} ] \right] \nonumber \\
        &= \E\left[ \frac{D}{\E[D]} C_t \right] - \E\left[ \frac{p(\mathcal{F}_{t^*-1})(1-D)} {\E[D](1-p(\mathcal{F}_{t^*-1}))} C_t \right] \label{eqn:thm-attc-2}
    \end{align}
    Further, notice that
    \begin{align}
        \E\left[ \frac{p(\mathcal{F}_{t^*-1})(1-D)} {\E[D](1-p(\mathcal{F}_{t^*-1}))} \right] &= \E\left[ \E\left[ \frac{p(\mathcal{F}_{t^*-1})(1-D)} {\E[D](1-p(\mathcal{F}_{t^*-1}))} \Big| \mathcal{F}_{t^*-1} \right] \right] \nonumber \\
        &= \frac{\E[ p(\mathcal{F}_{t^*-1}) ]}{\E[D]} \nonumber \\
        &= 1 \label{eqn:thm-attc-3}
    \end{align}
    Combining the results from \Cref{eqn:thm-attc-2,eqn:thm-attc-3} implies that
    \begin{align} \label{eqn:thm-attc-4}
        ATT^C_t = \E\left[ \omega(D,\mathcal{F}_{t^*-1}) C_t \right]
    \end{align}
    To conclude the proof, notice that
    \begin{align} \label{eqn:thm-attc-4b}
        \E[\omega(D,\mathcal{F}_{t^*-1}) | \mathcal{F}_{t^*-1}] &= \frac{1}{\E[D]}\Bigg( \E[D|\mathcal{F}_{t^*-1}] - \E\left[ \frac{p(\mathcal{F}_{t^*-1})(1-D)}{(1-p(\mathcal{F}_{t^*-1}))} \Big| \mathcal{F}_{t^*-1} \right]\Bigg) \nonumber \\
        &= 0
    \end{align}
    which holds from the definition of $\omega$ and also uses the argument in \Cref{eqn:thm-attc-3}.  This implies that
    \begin{align}
        \E\Big[\omega(D,\mathcal{F}_{t^*-1}) m_{0,t}^C(\mathcal{F}_{t^*-1}) \Big] &= \E\Big[ m_{0,t}^C(\mathcal{F}_{t^*-1}) \E[\omega(D,\mathcal{F}_{t^*-1}) | \mathcal{F}_{t^*-1}]  \Big] \nonumber \\
        &= 0 \label{eqn:thm-attc-5}
    \end{align}
    Combining \Cref{eqn:thm-attc-2} and \Cref{eqn:thm-attc-5} implies the second part of the result.
\end{proof}

\subsection*{Proof of \Cref{thm:atty}}

\begin{proof}
    For the first part, notice that
    \begin{align*}
        \E[\Delta^{(t^*-1,t)} I_t(0) | D=1] &= \E[I_t(0) | D=1] - \E[I_{t^*-1}|D=1] \\
        &= \E[I_t(1)|D=1] - (\E[I_t(1) | D=1] - \E[I_t(0) | D=1]) - \E[I_{t^*-1}|D=1] \\
        &= \E[\Delta^{(t^*-1,t)} I_t|D=1] - ATT_t^I \\
        &= \E\left[\frac{D}{\E[D]}\Delta^{(t^*-1,t)} I_t - \omega(D,\mathcal{F}_{t^*-1})(I_t - m_{0,t}^C(\mathcal{F}_{t^*-1})\right]
    \end{align*}
    where the first equality holds by splitting the difference, the second equality adds and subtracts $\E[I_t(1)|D=1]$, the third equality holds by combining terms and the definition of $ATT_t^I$, and the last equality holds by using the same arguments for $ATT_t^I$ as were used for $ATT_t^C$ in \Cref{thm:unc}.

    For the second part, to start with, notice that
    \begin{align}
        ATT^Y_t &= \E[Y_t(1) - Y_t(0) | D=1] \nonumber \\
        &= \E[Y_t(1) - Y_{t^*-1}(0) | D=1] - \E[Y_t(0) - Y_{t^*-1}(0) | D=1] \nonumber \\
        &= \E[\Delta^{(t^*-1,t)}Y_t | D=1] - \E[\Delta^{(t^*-1,t)} Y_t(0) | D=1] \label{eqn:thm3-atty}
    \end{align}
    where the first equality holds by the definition of $ATT^Y_t$, the second equality holds by adding and subtracting $\E[Y_{t^*-1}(0) | D=1]$, and the third equality holds by replacing the potential outcomes in the first term with their observed counterparts.  In \Cref{eqn:thm3-atty}, the first term is directly identified (it is the observed path of outcomes for treated locations) while the second term is not directly identified, and we consider it in detail below.  As a preliminary step, notice that
    \begin{align*}
        \E[\Delta^{(t^*-1,t)} Y_t(0) | D=0] = \tilde{\tau}_t + \alpha \E[\Delta^{(t^*-1,t)} I_{t}(0) | D=0]
    \end{align*}
    Since, $\Delta^{(t^*-1,t)} Y_{lt}(0)$ and $\Delta^{(t^*-1,t)} I_{lt}(0)$ are both observed outcomes for locations in the untreated group, this implies that $\tilde{\tau}_t$ and $\alpha$ are identified and can be recovered from the regression of $\Delta^{(t^*-1,t)} Y_{lt}$ on $\Delta^{(t^*-1,t)} I_{lt}$ among the untreated group. Next, 
    \begin{align*}
        \E[\Delta^{(t^*-1,t)} Y_t(0) | D=1] &= \tilde{\tau}_t + \alpha \E[\Delta^{(t^*-1,t)} I_{t}(0) | D=1] \\
        &= \tilde{\tau}_t + \alpha\left(\E\left[\frac{D}{\E[D]}\Delta^{(t^*-1,t)} I_t - \omega(D,\mathcal{F}_{t^*-1})(I_t - m_{0,t}^C(\mathcal{F}_{t^*-1})\right] \right)
    \end{align*}
    and all terms are identified.  Plugging this expression into \Cref{eqn:thm3-atty} implies the result.
    
    Next, we consider the bias coming from using standard DID.  This is given by
    \begin{align*}
        & \E[\Delta^{(t^*-1,t)} Y_t | D=1] - \E[\Delta^{(t^*-1,t)} Y_t | D=0] - \big( \E[Y_t(1) - Y_t(0) | D=1])\\
        & \hspace{50pt} = \E[\Delta^{(t^*-1,t)} Y_t(0) | D=1] - \E[\Delta^{(t^*-1,t)} Y_t(0) | D=0] \\
        & \hspace{50pt} =\alpha \big(\E[\Delta^{(t^*-1,t)} I_t(0) | D=1] - \E[\Delta^{(t^*-1,t)} I_t(0) | D=0] \big)
    \end{align*}
    where the second equality holds by adding and subtracting $\E[Y_{t^*-1}(0) | D=1]$ and canceling terms and the last equality holds by the model in \Cref{eqn:economic-outcomes-twfe}.
    
    Finally, we consider the bias coming from DID including actual cases (rather than cases in the absence of the policy) as a covariate.  This bias is given by
    \begin{align*}
        & \E[\Delta^{(t^*-1,t)} Y_t | D=1] - \big(\tilde{\tau}_t + \alpha \E[\Delta^{(t^*-1,t)} I_t | D=1]\big) - \big(\E[Y_t(1)-Y_t(0) | D=1]\big) \\
        & \hspace{50pt} = \alpha \big( \E[\Delta^{(t^*-1,t)} I_t(0) | D=1] - \E[\Delta^{(t^*-1,t)} I_t | D=1] \big) \\
        & \hspace{50pt} = - \alpha ATT^I_t
    \end{align*}
    where the first equality holds by adding and subtracting $\E[Y_{t^*-1}(0)|D=1]$ and from the model in \Cref{eqn:economic-outcomes-twfe} (and by canceling terms), and the last equality holds by the definition of $ATT^I_t$ after canceling terms.
    
\end{proof}

\end{document}